\documentclass[a4paper,twoside]{article}
\usepackage{tikz}
\usepackage[utf8]{inputenc}
\usepackage[ukrainian,english]{babel}
\usepackage{subcaption} 

\usepackage{eejp}

\begin{document}

\title[Benjamin–-Feir Instability of in a Two-Layer Fluid. Part I]
{Benjamin–-Feir Instability of Interfacial Gravity–Capillary Waves in a Two-Layer Fluid. Part I}

\author{0000-0002-7960-1436}{Olga Avramenko}{a*,b}
\author{0000-0001-5187-8831}{Volodymyr Naradovyi}{c}
\authors[O. Avramenko, V. Naradovyi]%
{Olga Avramenko, Volodymyr Naradovyi}


\BeginPaper 
\begin{center}
\AuthorPrint 
\affiliation{a}{National University of Kyiv-Mohyla Academy, 2 Skovorody,
        Kyiv, 04070, Ukraine}
\affiliation{b}{Vytautas Magnus University, 58 K.~Donelaičio g.,
        Kaunas 44248, Lithuania}
\affiliation{c}{Volodymyr Vynnychenko Central Ukraine State University, 1 Shevchenka, Kropyvnytskyi, 25006, Ukraine}
\email{o.avramenko@ukma.edu.ua}

\data{Received Month XX, 20XX; revised Month XX, 20XX; accepted December Month XX, 20XX}
\end{center}



\newcommand{\ep}{\varepsilon}
\newcommand{\eps}[1]{{#1}_{\varepsilon}}



\begin{abstract}
This study presents a detailed investigation of the modulational stability of interfacial wave packets in a two-layer inviscid incompressible fluid with finite layer thicknesses and interfacial surface tension. The stability analysis is carried out for a broad range of density ratios and geometric configurations, enabling the construction of stability diagrams in the $(\rho,k)$-plane, where $\rho$ is the density ratio and $k$ is the carrier wavenumber. The Benjamin-–Feir index is used as the stability criterion, and its interplay with the curvature of the dispersion relation is examined to determine the onset of modulational instability.
The topology of the stability diagrams reveals several characteristic structures: a localized \emph{loop} of stability within an instability zone, a global \emph{upper} stability domain, an elongated \emph{corridor} bounded by resonance and dispersion curves, and a degenerate \emph{cut} structure arising in strongly asymmetric configurations. Each of these structures is associated with a distinct physical mechanism involving the balance between focusing/defocusing nonlinearity and anomalous/normal dispersion.
Systematic variation of layer thicknesses allows us to track the formation, deformation, and disappearance of these regions, as well as their merging or segmentation due to resonance effects. Limiting cases of semi-infinite layers are analyzed to connect the results with known configurations, including the `half-space–layer', `layer–half-space', and `half-space–half-space' systems. The influence of symmetry and asymmetry in layer geometry is examined in detail, showing how it governs the arrangement and connectivity of stable and unstable regions in parameter space.
The results provide a unified framework for interpreting modulational stability in layered fluids with interfacial tension, highlighting both global dispersion-controlled regimes and localized stability islands. This work constitutes Part~I of the study; Part~II will address the role of varying surface tension, which is expected to deform existing stability domains and modify the associated nonlinear–dispersive mechanisms.

\key{modulational stability, interfacial waves, two-layer fluid, Benjamin-–Feir instability, surface tension} 

\end{abstract}


\section{Introduction}
\label{sec_0}

This study investigates wave packet propagation along a two-layer fluid interface with surface tension, focusing on modulational instability as a fundamental nonlinear mechanism. First described by Benjamin~\cite{Benjamin1967} and formulated via the nonlinear Schr\"{o}dinger (NLS) equation by Zakharov~\cite{Zakharov1968}, modulational instability plays a central role in nonlinear wave evolution. Classical multiple-scale derivations of the NLS equation for modulated wave packets in layered fluids were carried out by Hasimoto and Ono~\cite{Hasimoto1972} for a single layer and by Nayfeh~\cite{Nayfeh1976} for a two-layer configuration, with later refinements for interfacial waves by Grimshaw and Pullin~\cite{Grimshaw85} and Selezov et~al.~\cite{Selezov03}. These works established the theoretical basis for studying stability boundaries in terms of nonlinear–dispersive balances; however, they did not include the specific `layer with a solid bottom – layer with a rigid lid' (La–La) configuration considered here.

A broad range of studies has addressed nonlinear wave propagation in two-layer and interfacial systems. Oikawa~\cite{Oikawa89} analyzed resonant interactions between internal gravity waves and surface wave packets, while Ablowitz and Haut~\cite{Ablowitz09} developed coupled NLS models for interfacial flows with a free surface. Christodoulides and Dias~\cite{Christodoulides1995} investigated the stability of capillary–gravity interfacial waves between two bounded fluid layers, including limiting configurations such as `lower half-space–upper layer with rigid lid' and `half-space–half-space'. Avramenko et~al.~\cite{Avramenko16} examined dispersion relations in a two-layer liquid with a finite-depth free surface, and Panda and Martha~\cite{Panda17} incorporated surface tension in scattering problems over a permeable bottom. Purkait and Debsarma~\cite{Purkait_Debsarma_2019} studied oblique wave packet interactions in a two-layer fluid with an infinitely deep lower layer, finding that instability growth rates depend strongly on upper layer thickness. More recent works by Li et~\cite{Li_Cao_Song_Yu_Chen_2020,Li_Song_Cao_2020,Li19,Li22} and Pal and Dhar~\cite{Pal22,Pal24} extended the multiple-scale framework to include arbitrary depths, surface tension, and background currents, revealing significant modifications of instability thresholds and growth rates. While these studies identify key physical mechanisms in specific configurations, none systematically characterizes modulational stability in a La–La system with arbitrary finite thicknesses.

Benjamin--Feir type instability of internal gravity waves has also been widely investigated in geophysical and atmospheric contexts, revealing diverse instability mechanisms. Talipova et~al.~\cite{Talipova2011} demonstrated modulational instability of long internal waves in horizontally inhomogeneous oceans, while Liang et~al.~\cite{Liang2016} identified inherent instability via resonant harmonic generation. Chow et~al.~\cite{Chow2019} analyzed instability in smoothly stratified shallow fluids, and Voelker and Schlutow~\cite{Voelker2023} reported cascades of modulation and shear instabilities in vertically sheared flows. Kuklin and Poklonskiy~\cite{Kuklin2019} examined anomalous amplitude waves in the ocean, showing their persistence, enhanced modulation instability, and ability to propagate over long distances. Recent rigorous analyses, such as Bianchini et~al.~\cite{Bianchini2025} for the Boussinesq system and Lashkin and Cheremnykh~\cite{Lashkin2024} for atmospheric waves, provide quantitative thresholds for instability onset.
Although these studies deepen understanding of modulational instability in complex media, the combined effects of finite thickness and surface tension in the La--La configuration remain unexplored.

There has also been sustained interest in the role of surface tension and higher-order effects in wave stability. Ionescu and Pusateri~\cite{Ionescu2018} and D\"{u}ll~\cite{Dull2021} examined global regularity and the validity of the NLS approximation for water waves with capillarity, while Sedletsky~\cite{Sedletsky21} derived a fifth-order NLS equation for finite-depth gravity–capillary waves. Ablowitz et~al.~\cite{Ablowitz23} analyzed six-wave interactions and resonances, demonstrating the complexity of nonlinear–dispersive interplay with surface tension. However, no comprehensive topological description of modulational stability diagrams has been given for a two-layer finite-depth La–La system.

The present paper (Part~I) addresses this gap using multiscale expansion with symbolic computation, verifying results through limiting transitions to the `half-space–layer' (HS–La), `layer–half-space' (La–HS), and `half-space–half-space' (HS–HS) configurations. A companion study (Part~II) will focus on the influence of surface tension magnitude on the topology of stability boundaries, revealing additional stabilization and destabilization mechanisms.

\section{Problem formulation and some preliminary background}
\label{KVSec_1}

\subsection{Wave packets in a two-layer fluid and their modulational stability condition}

\label{SubSec_Statement}

We consider the propagation of wave packets along the interface
\( z = \eta(x, t) \) between two incompressible fluid layers,
\( \Omega_{1} \) and \( \Omega_{2} \), with densities
\( \rho_{1} \) and \( \rho_{2} \), respectively.
The effect of interfacial surface tension \( T \) acting on
\( \eta(x,t) \) is taken into account.
In the undisturbed state, the domains are $\Omega_{1} = \{(x,z) \,:\, |x| < \infty,\ - h_{1} < z < 0\}, \quad \Omega_{2} = \{(x,z) \,:\, |x| < \infty,\ 0 < z < h_2\}, $
where \(h_{1}\) and \(h_{2}\) denote the layer thicknesses.

The problem is formulated in dimensionless variables by introducing characteristic scales based on gravitational acceleration \( g \), the density of the lower fluid \( \rho_{1} \), and a reference surface tension \( T_{0} \). The characteristic length, time, and mass are defined as \( L = (T_{0}/(\rho_{1} g))^{1/2} \), \( t_{0} = (L/g)^{1/2} \), and \( m_{0} = \rho_{1} L^{3} \). Dimensionless variables (denoted by an asterisk) are then introduced as \((x^{*},z^{*},h_{1}^{*})=(x,z,h_{1})/L\), \((\rho_{1}^{*},\rho_{2}^{*})=(\rho_{1},\rho_{2})/\rho_{1}\), \(t^{*}=t/t_{0}\), \(T^{*}=T/T_{0}\), \(\eta^{*}=\eta/(\alpha L)\), and \((\phi_{1}^{*},\phi_{2}^{*})=(\phi_{1},\phi_{2})/(\alpha L^{2}/t_{0})\), where \(\alpha=a/l\) is the small wave–steepness parameter, with \(a\) the maximum interface displacement and \(l\) the wavelength, and \(\rho=\rho_{2}/\rho_{1}\) is the density ratio. In the analytical derivations the parameter \(T\) is retained explicitly, while in the numerical analysis and in all stability diagrams presented below the dimensionless surface tension coefficient is fixed at \(T=1\).

In nondimensional variables, the problem is formulated in terms of the velocity potentials \(\phi_{j}(x,z,t)\) and the interface displacement \(\eta(x,t)\), with the governing equations and boundary conditions given by
\begin{align}
   &\Delta \phi_{j} = 0, \qquad (x,z) \in \Omega_{j}, \nonumber \\
   &\eta_{,t} - \phi_{j,z} = - \alpha\, \eta_{,x} \,\phi_{j,x},
       \qquad z = \alpha \eta(x,t), \nonumber \\
   &\phi_{1,t}  -  \rho \phi_{2,t} + (1 - \rho)\,\eta
       + \tfrac12 \alpha  \left( \nabla \phi_{1} \right)^{2}
       - \tfrac12 \alpha \rho  \left( \nabla \phi_{2} \right)^{2}
    - T  \left[ 1 + \left( \alpha \eta_{,x} \right)^{2} \right]^{-3/2}
       \eta_{,xx} = 0,
       \qquad z = \alpha \eta(x,t), \nonumber \\
   &\phi_{1,z} = 0, \quad z = -h_{1}, \qquad\qquad
       \phi_{2,z} = 0, \quad z = h_{2}.
\label{eq:Statement}
\end{align}
Following the multiple-scale method, we represent the unknown fields as
\begin{equation}
    (\eta, \phi_j) =
    \sum_{n=1}^{3} \alpha^{n-1} \left( \eta_{n}, \phi_{jn} \right)
    + O(\alpha^{3}),
    \nonumber
    \label{eq:expansion}
\end{equation}
with slow variables \(x_{n}=\alpha^{n}x\), \(t_{n}=\alpha^{n}t\) (\(n=0,1,2\)), and expansion coefficients \(\eta_{n}\), \(\phi_{jn}\).

From the first approximation of the problem (\ref{eq:Statement}) we obtain the dispersion relation
\begin{equation}
\omega^2 = \frac{k \left( 1 - \rho + T k^2 \right)}
               {\coth k h_1 + \rho\,\coth k h_2},
\label{eq:dispersion}
\end{equation}
where \(k\) is the carrier wave number and \(\omega\) the carrier frequency.

The second-order problem yields the amplitude of the second harmonic in the form
\begin{align}
\Lambda &= 0.5 k \omega^2  \frac{\rho\,\coth^2 k h_2 - \coth^2 k h_1
    + 4 \rho\,\coth 2k h_2\,\coth k h_2- 4 \coth 2k h_1\,\coth k h_1
    - 3(\rho - 1)}
 {4 T k^3 - k \rho + k
    - 2 \omega^2 \left( \rho\,\coth 2k h_2
    + \coth 2k h_1 \right)}.
\label{eq:Lambda}
\end{align}

At third order, the solvability condition together with
\eqref{eq:dispersion} and \eqref{eq:Lambda} gives the nonlinear Schr\"{o}dinger equation
\begin{equation}
    i A_{,t} + i \omega^{\prime} A_{,x}
    + 0.5 \omega^{\prime \prime} A_{,xx}
    = - \alpha\, \omega^{-1} J\, A^{2} \bar{A},
    \label{eq:NLS}
\end{equation}
where \(\overline{A}\) is the complex conjugate of \(A\),
\(\omega^{\prime} = \partial \omega / \partial k\) is the group velocity,
and \(\omega^{\prime \prime} = \partial^{2} \omega / \partial k^{2}\).

The Benjamin–Feir index \(J\) is explicitly
\begin{align}
J &= - \frac{1}{16(1 - \rho) \left[ \rho\,\coth k h_2 + \coth k h_1 \right]} \bigg\{
    2 k \omega^2 (1 - \rho) \Lambda \left[ -3 \rho\,\coth^2 k h_2
      + 3 \coth^2 k h_1 - 1 + \rho \right] \nonumber\\
&\qquad - 4 k \omega^4 \left[ \rho \left( \coth^2 k h_2 - 1 \right)
      - \left( \coth^2 k h_1 - 1 \right) \right]^2  \nonumber\\
&\qquad - 4 k^2 \omega^2 (1 - \rho) \left[ \rho\,\coth^3 k h_2
      + \coth^3 k h_1 - 2 \rho\,\coth k h_2 - 2 \coth k h_1 \right]   - 3 T k^5 (1 - \rho)\bigg\}.
\label{eq:J}
\end{align}
A purely temporal solution of \eqref{eq:NLS} reads
\(
A(t) = a \,\exp\!\left( i \alpha a^2 \omega^{-1} J\, t \right),
\)
where \(a\) is a constant envelope amplitude.
Following \cite{Nayfeh1976}, the modulational stability condition is
obtained by linearizing equation~\eqref{eq:NLS} around a uniform
envelope solution and analysing the evolution of small harmonic
sideband perturbations. The sign of the product
\(J\,\omega^{\prime\prime}\) determines the outcome: stability occurs
when
\begin{equation}
J\,\omega^{\prime\prime} < 0,
\label{eq:mod_stability}
\end{equation}
while \(J\,\omega^{\prime\prime} > 0\) indicates modulational
instability.

The criterion \eqref{eq:mod_stability} provides a simple yet
powerful diagnostic for determining whether a quasi-monochromatic
wave packet is modulationally stable or unstable.
If \(J\,\omega^{\prime\prime} < 0\), the effects of nonlinearity and
dispersion counteract each other, preventing the growth of sideband
perturbations and ensuring the persistence of a uniform packet.
Conversely, when \(J\,\omega^{\prime\prime} > 0\), nonlinearity and
dispersion act in unison, amplifying small sideband disturbances.
This cooperative interaction leads to the exponential growth of
modulation, breakdown of the initial packet, and, in many cases,
the emergence of localized structures such as envelope solitons or
wave focusing events. The sign of \(J\) is determined by the balance
between focusing/defocusing nonlinearity, while \(\omega^{\prime\prime}\)
reflects the curvature of the dispersion relation and thus the type
of dispersion (normal or anomalous) present in the system.

\subsection{Limiting cases}
\label{SubSec_LimitingCases}

The La--La system with finite layer thicknesses $h_1$ and $h_2$ reduces in several limiting cases to previously studied configurations:
(i) $h_1 \to \infty$ $\Rightarrow$ HS--La \cite{AvramenkoNarad2025};
(ii) $h_2 \to \infty$ $\Rightarrow$ La--HS \cite{AvramenkoNarad2026};
(iii) $h_1,h_2 \to \infty$ $\Rightarrow$ HS--HS \cite{Nayfeh1976}.
All key relations---the dispersion relation \eqref{eq:dispersion}, the second-harmonic amplitude \eqref{eq:Lambda}, the evolution equation \eqref{eq:NLS}, the Benjamin--Feir index \eqref{eq:J}, and the modulational stability condition \eqref{eq:mod_stability}---degenerate to their known forms in these limits.

In agreement with \cite{Nayfeh1976, Christodoulides1995, AvramenkoNarad2026}, the condition $J=0$ as $k \to \infty$ yields two vertical asymptotes at
\begin{equation} \label{eq:asymptotes}
\rho = \frac{2 - \sqrt{2}}{2 + \sqrt{2}} \simeq 0.1716,
\qquad
\rho = \frac{2 + \sqrt{2}}{2 - \sqrt{2}} \simeq 5.8275.
\end{equation}

In the special case $\rho = 1$ (equal densities), the gravitational restoring mechanism vanishes, but for $T>0$ the interface still supports waves due solely to surface tension. This regime models a homogeneous medium with a purely capillary interface, exhibiting nontrivial dispersion distinct from both surface and internal waves.

The limit $\rho \to 1$ of $J$ is
\begin{equation}\label{J_lim}
\lim_{\rho\to 1} J =
\begin{cases}
\dfrac{4k \coth^3 kh - \coth^2 kh - 5k \coth kh + 1}
{8\coth^2 kh}\,T\, k^4, & h_1=h_2=h,\\[1ex]
-\infty, & h_1\neq h_2.
\end{cases}
\end{equation}
For $h_1\neq h_2$, $J\to -\infty$ for all $k$, making $\rho=1$ a universal boundary between stable and unstable regimes. For $h_1=h_2$, the limit is finite and sign-changing with $k$, so $\rho=1$ does not universally separate stability and instability; the outcome depends on local spectral properties.

\section{Analysis of the evolution of modulational stability diagrams with varying layer thicknesses}

\subsection{Notation overview for stability diagrams} \label{sec:notation}

This subsection provides a general description of the notation and principles used in constructing modulational stability diagrams in the $(\rho,k)$ parameter plane for a La--La system.
Each diagram contains two principal regions: the zones of linear instability, indicated by a dark shade, and the zones of linear stability.
The latter are further subdivided into subregions of nonlinear stability, shown as unshaded areas, and modulational instability (i.e., instability of the wave packet envelope or Benjamin--Feir instability), displayed in a light shade.

The boundary between linear stability and instability is determined by the condition
\(
k > k_{c} = \sqrt{(\rho - 1)/T},
\)
and the curve $k = k_{c}$, drawn in black, divides the $(\rho,k)$ plane into a region of linear instability (below this curve) and a region of linear stability (above and to the left).
Within the linear stability region, the modulational behaviour is determined by the sign of the product $J\omega''$.
The subdivision into zones is carried out by the curves $J = 0$ (red), $J \to \infty$ (blue), and $\omega'' = 0$ (green), corresponding respectively to changes in the sign of the nonlinearity, resonant conditions, and changes in the sign of the second derivative of the carrier frequency $\omega''$ with respect to the wavenumber $k$, or, equivalently, of the group velocity.

\begin{figure}
\centering

\hspace{-5ex}
\begin{subfigure}[b]{0.29\textwidth}
  \includegraphics[width=\linewidth]{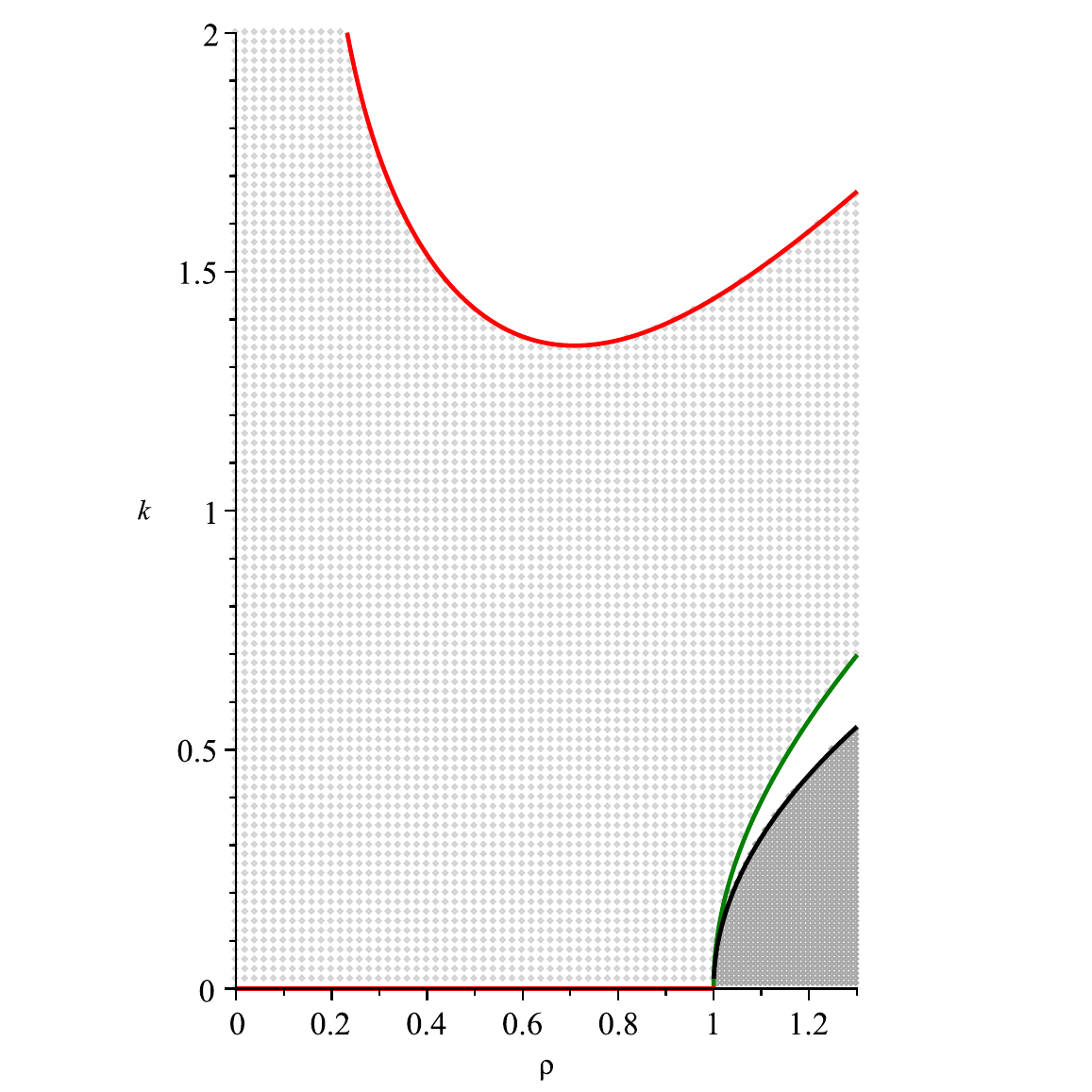}
  \caption{$h_1\!=\!1,\, h_2\!=\!1$}
  \label{fig:Fig1a}
\end{subfigure}\hspace{-6ex}
\begin{subfigure}[b]{0.29\textwidth}
  \includegraphics[width=\linewidth]{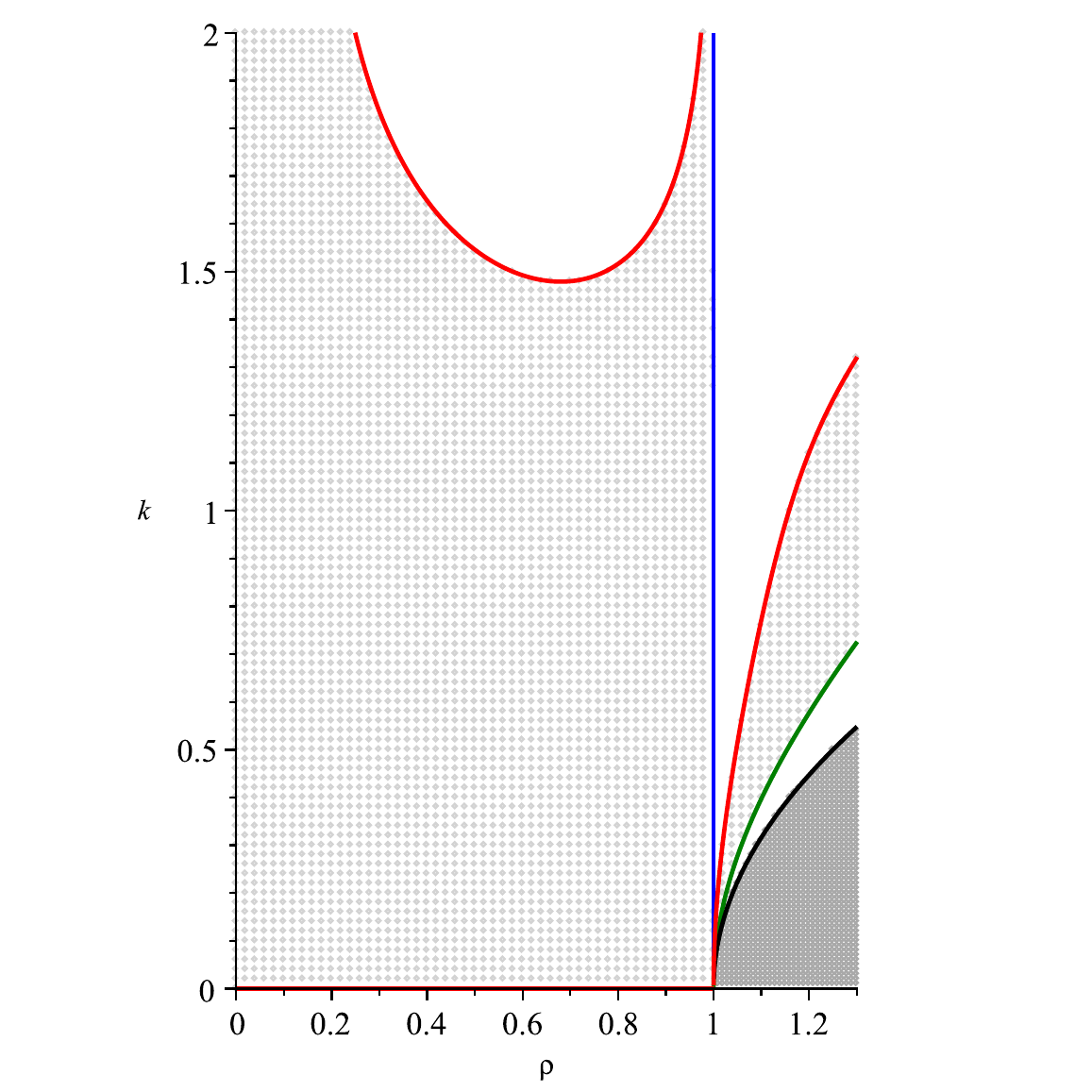}
  \caption{$h_1\!=\!1,\, h_2\!=\!2$}
  \label{fig:Fig1b}
\end{subfigure}\hspace{-6ex}
\begin{subfigure}[b]{0.29\textwidth}
  \includegraphics[width=\linewidth]{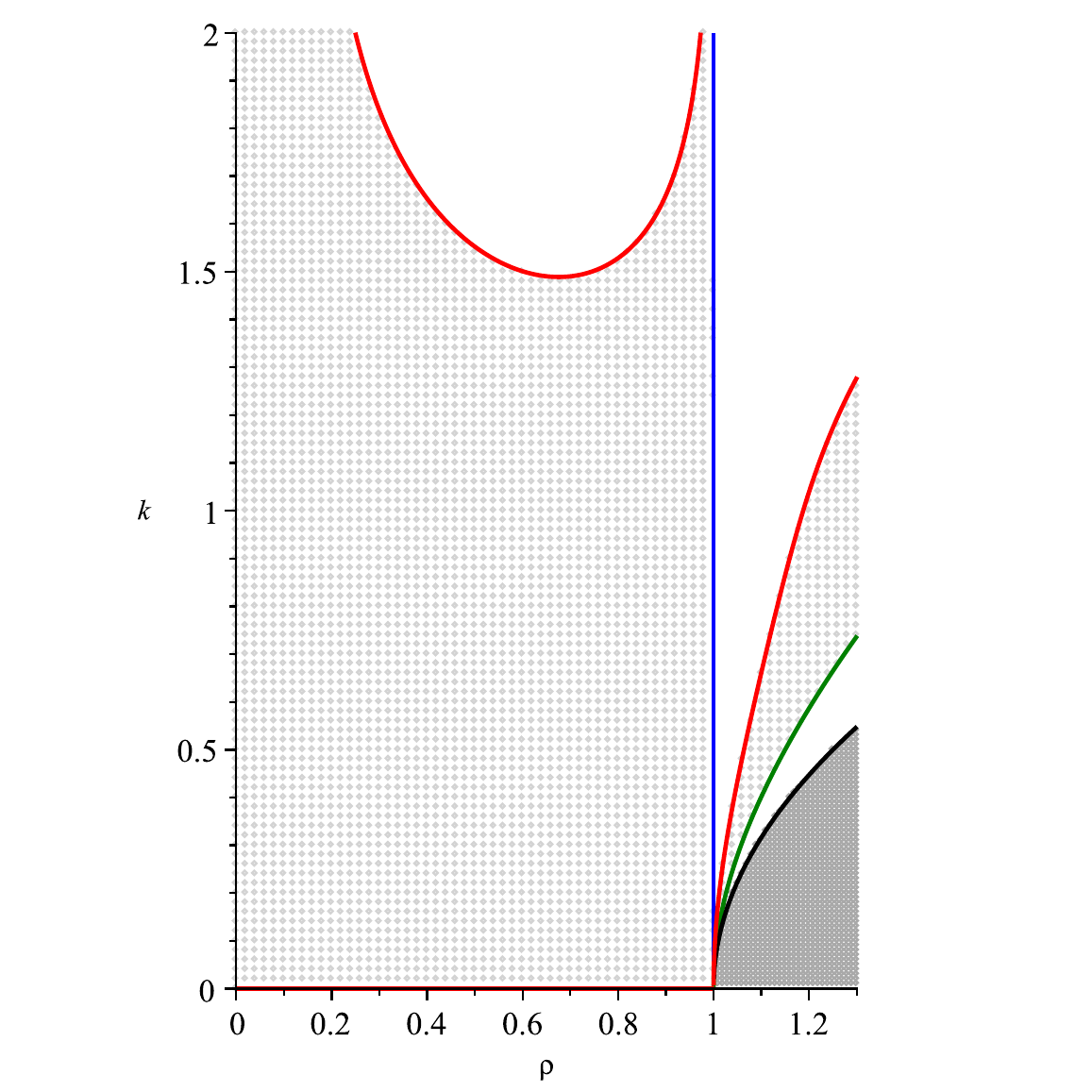}
  \caption{$h_1\!=\!1,\, h_2\!=\!3$}
  \label{fig:Fig1c}
\end{subfigure}\hspace{-6ex}
\begin{subfigure}[b]{0.29\textwidth}
  \includegraphics[width=\linewidth]{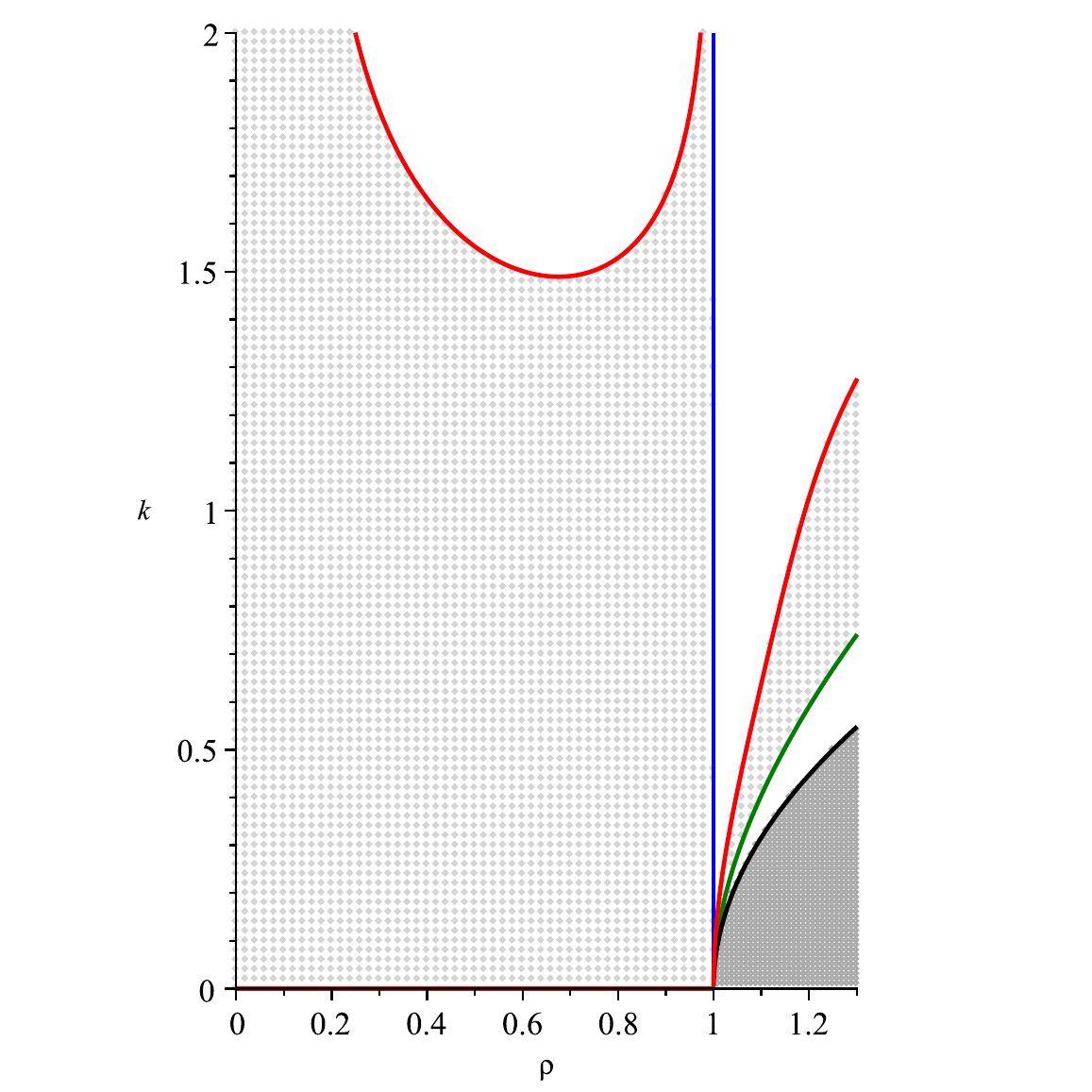}
  \caption{$h_1\!=\!1,\, h_2\!=\!4$}
  \label{fig:Fig1d}
\end{subfigure}

\vspace{0.5em}

\hspace{-5ex}
\begin{subfigure}[b]{0.29\textwidth}
  \includegraphics[width=\linewidth]{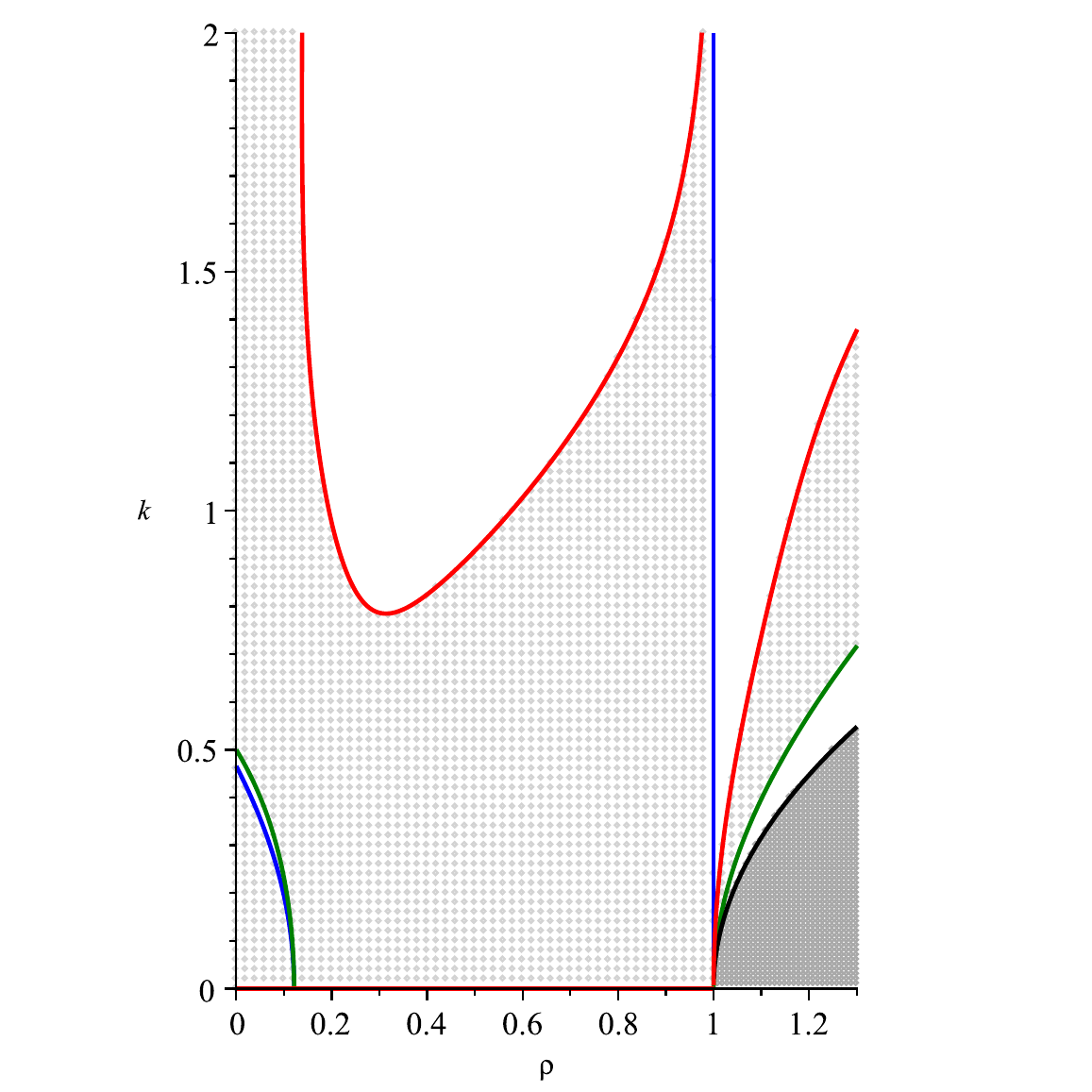}
  \caption{$h_1\!=\!2,\, h_2\!=\!1$}
  \label{fig:Fig1e}
\end{subfigure}\hspace{-6ex}
\begin{subfigure}[b]{0.29\textwidth}
  \includegraphics[width=\linewidth]{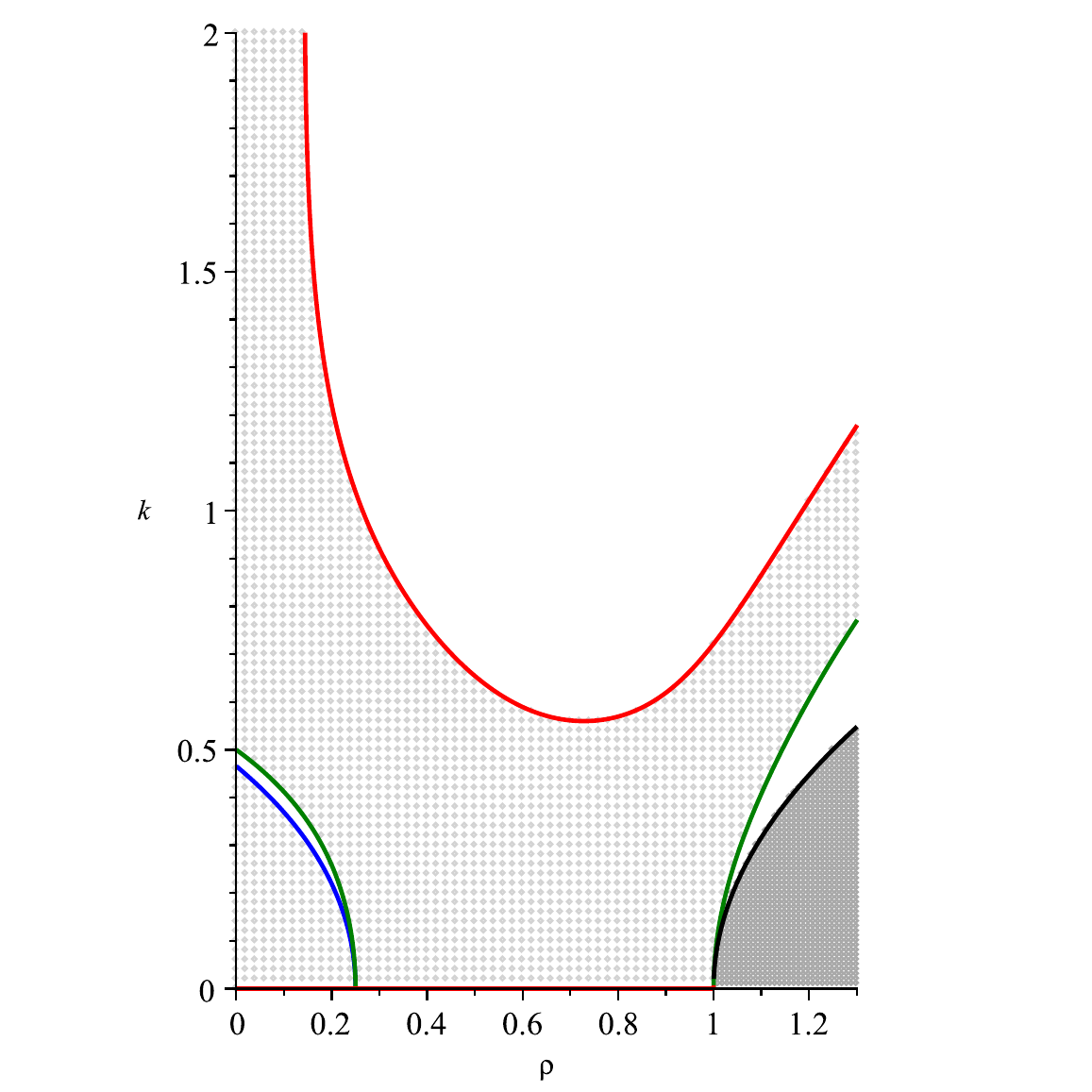}
  \caption{$h_1\!=\!2,\, h_2\!=\!2$}
  \label{fig:Fig1f}
\end{subfigure}\hspace{-6ex}
\begin{subfigure}[b]{0.29\textwidth}
  \includegraphics[width=\linewidth]{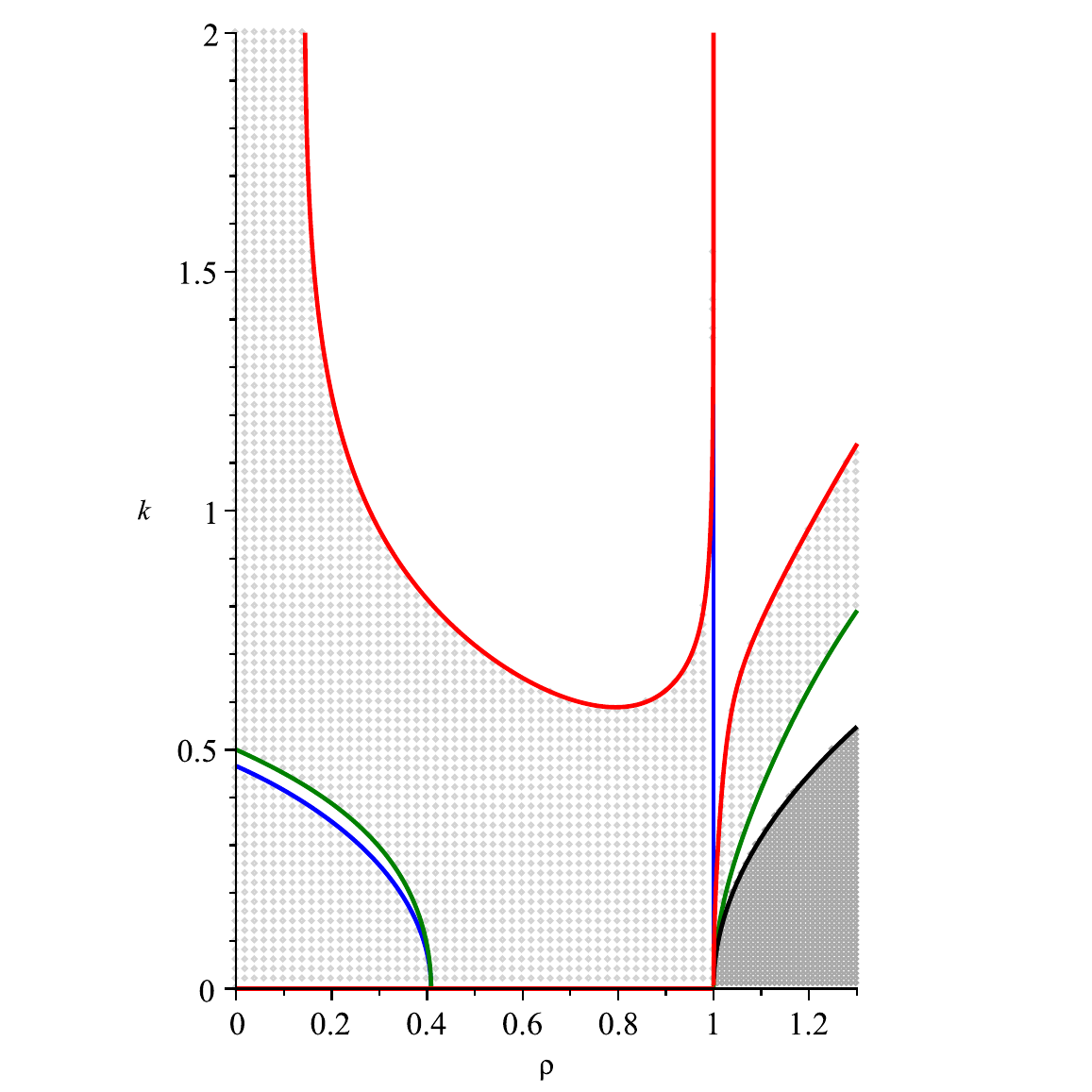}
  \caption{$h_1\!=\!2,\, h_2\!=\!3$}
  \label{fig:Fig1g}
\end{subfigure}\hspace{-6ex}
\begin{subfigure}[b]{0.29\textwidth}
  \includegraphics[width=\linewidth]{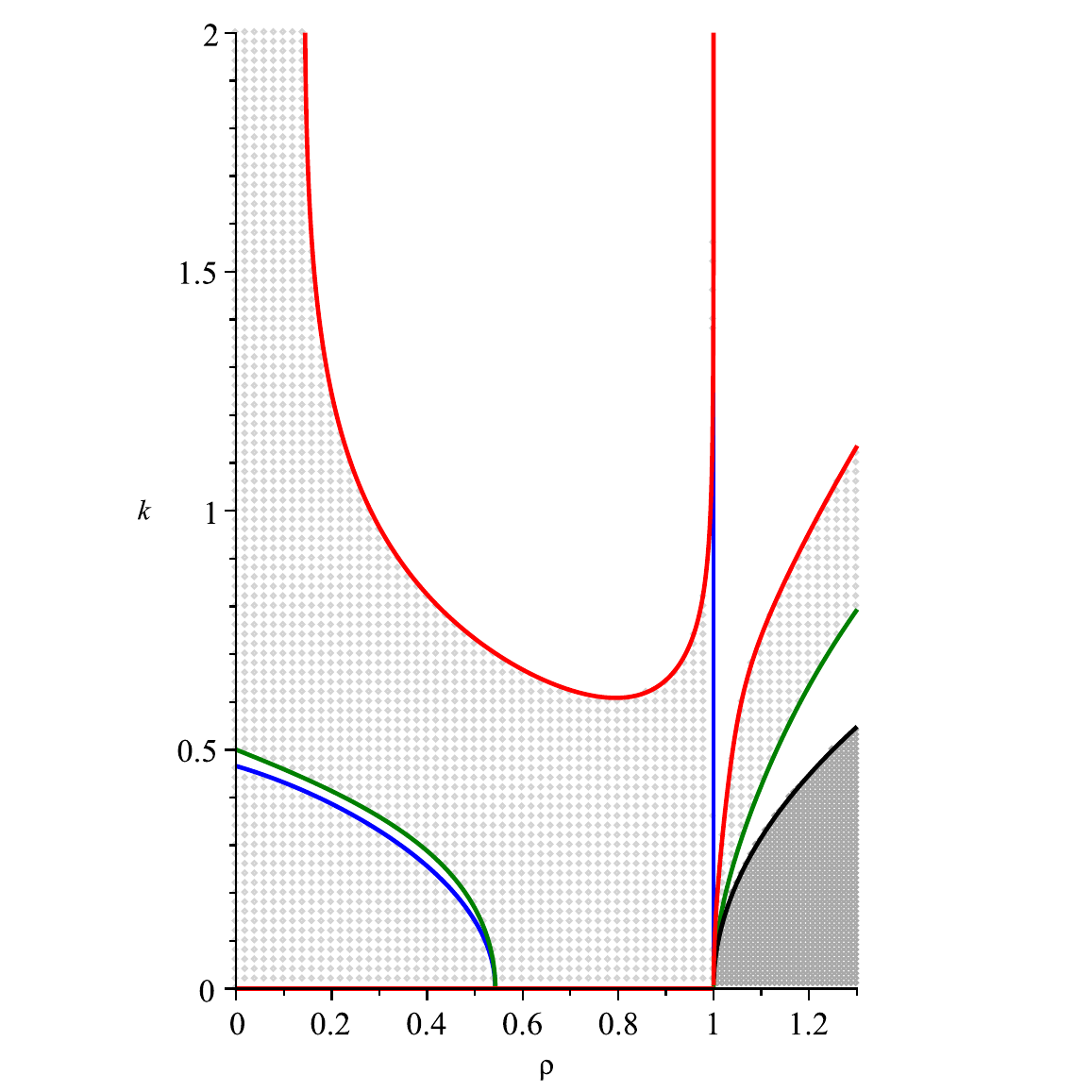}
  \caption{$h_1\!=\!2,\, h_2\!=\!4$}
  \label{fig:Fig1h}
\end{subfigure}

\vspace{0.5em}

\hspace{-5ex}
\begin{subfigure}[b]{0.29\textwidth}
  \includegraphics[width=\linewidth]{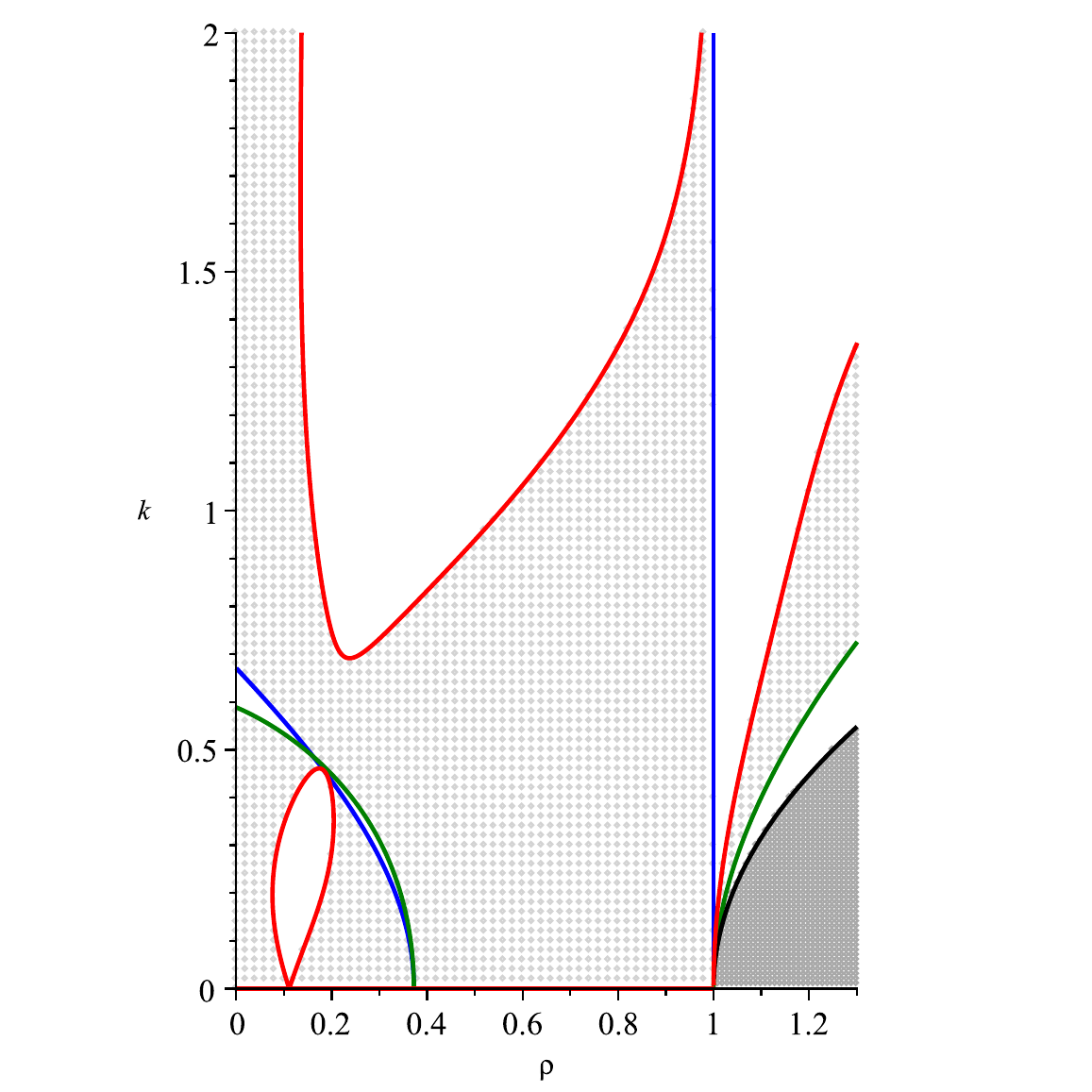}
  \caption{$h_1\!=\!3,\, h_2\!=\!1$}
  \label{fig:Fig1i}
\end{subfigure}\hspace{-6ex}
\begin{subfigure}[b]{0.29\textwidth}
  \includegraphics[width=\linewidth]{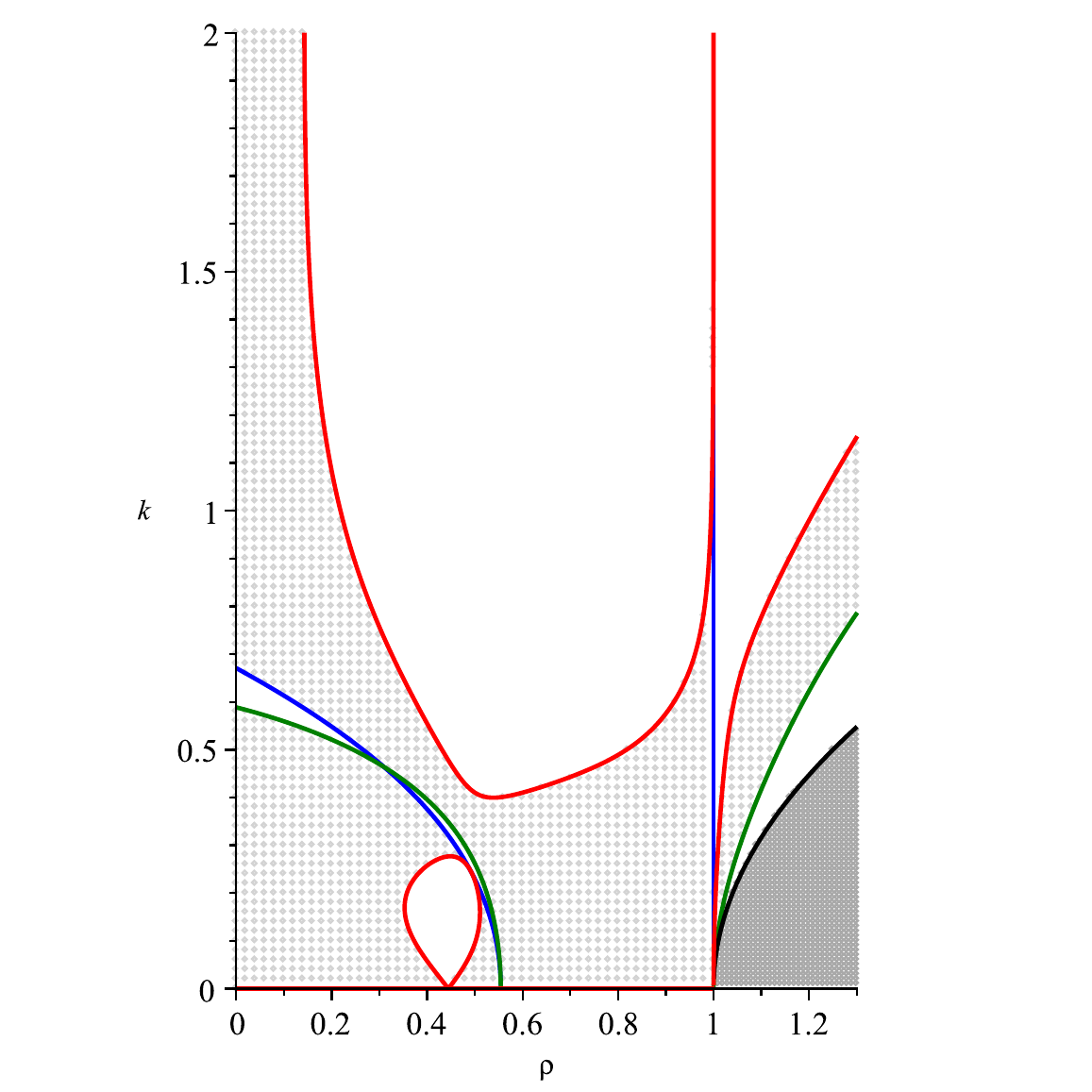}
  \caption{$h_1\!=\!3,\, h_2\!=\!2$}
  \label{fig:Fig1j}
\end{subfigure}\hspace{-6ex}
\begin{subfigure}[b]{0.29\textwidth}
  \includegraphics[width=\linewidth]{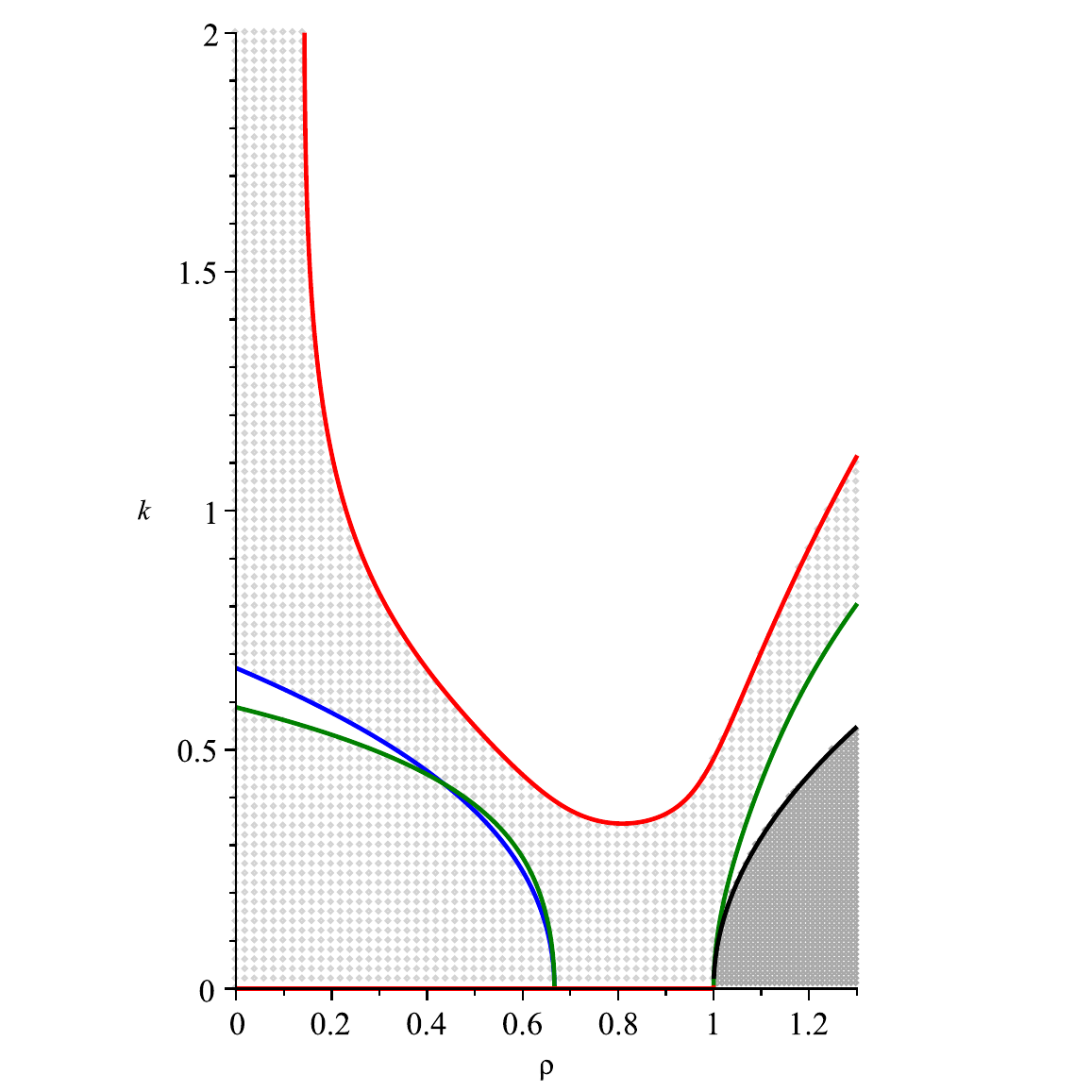}
  \caption{$h_1\!=\!3,\, h_2\!=\!3$}
  \label{fig:Fig1k}
\end{subfigure}\hspace{-6ex}
\begin{subfigure}[b]{0.29\textwidth}
  \includegraphics[width=\linewidth]{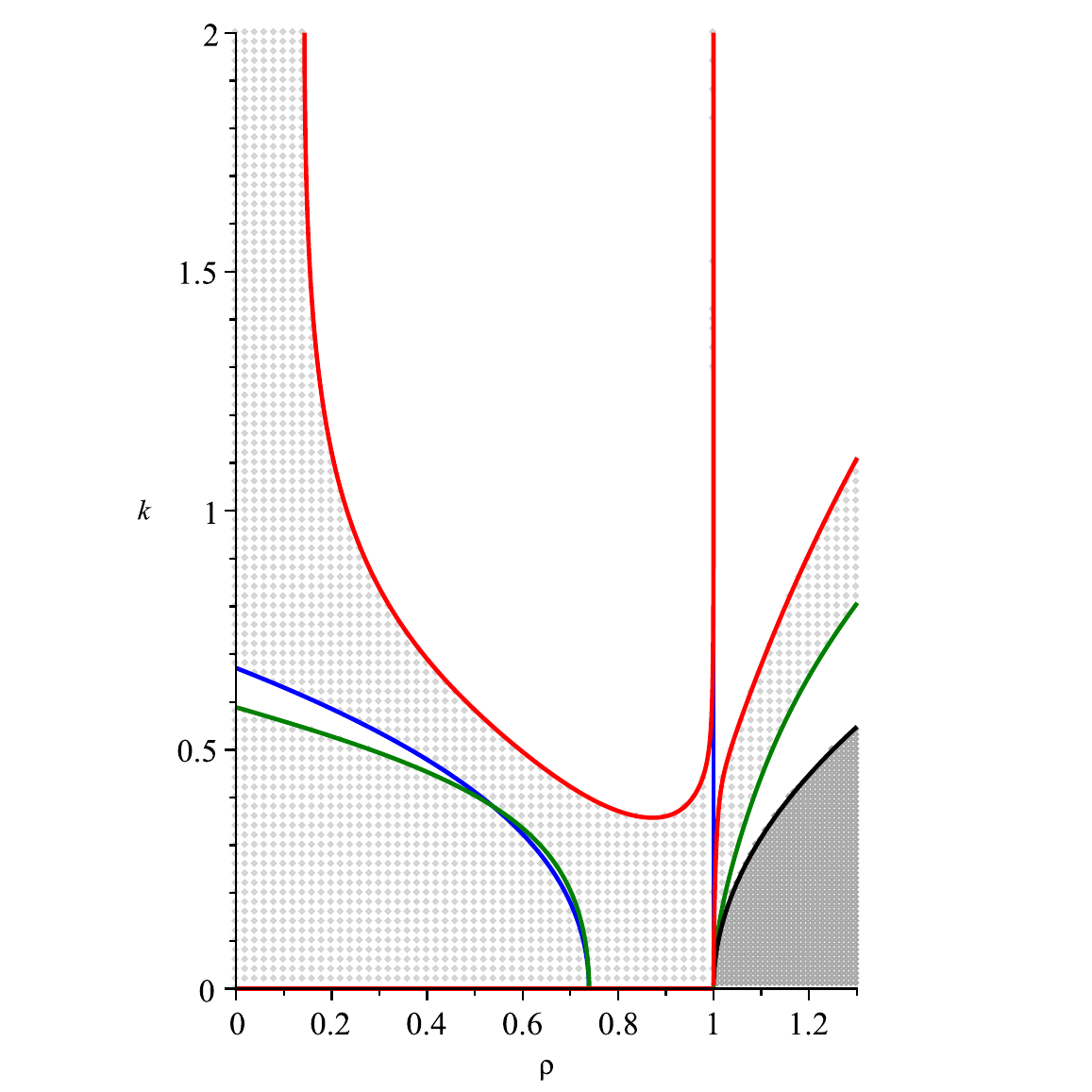}
  \caption{$h_1\!=\!3,\, h_2\!=\!4$}
  \label{fig:Fig1l}
\end{subfigure}

\vspace{0.5em}

\hspace{-5ex}
\begin{subfigure}[b]{0.29\textwidth}
  \includegraphics[width=\linewidth]{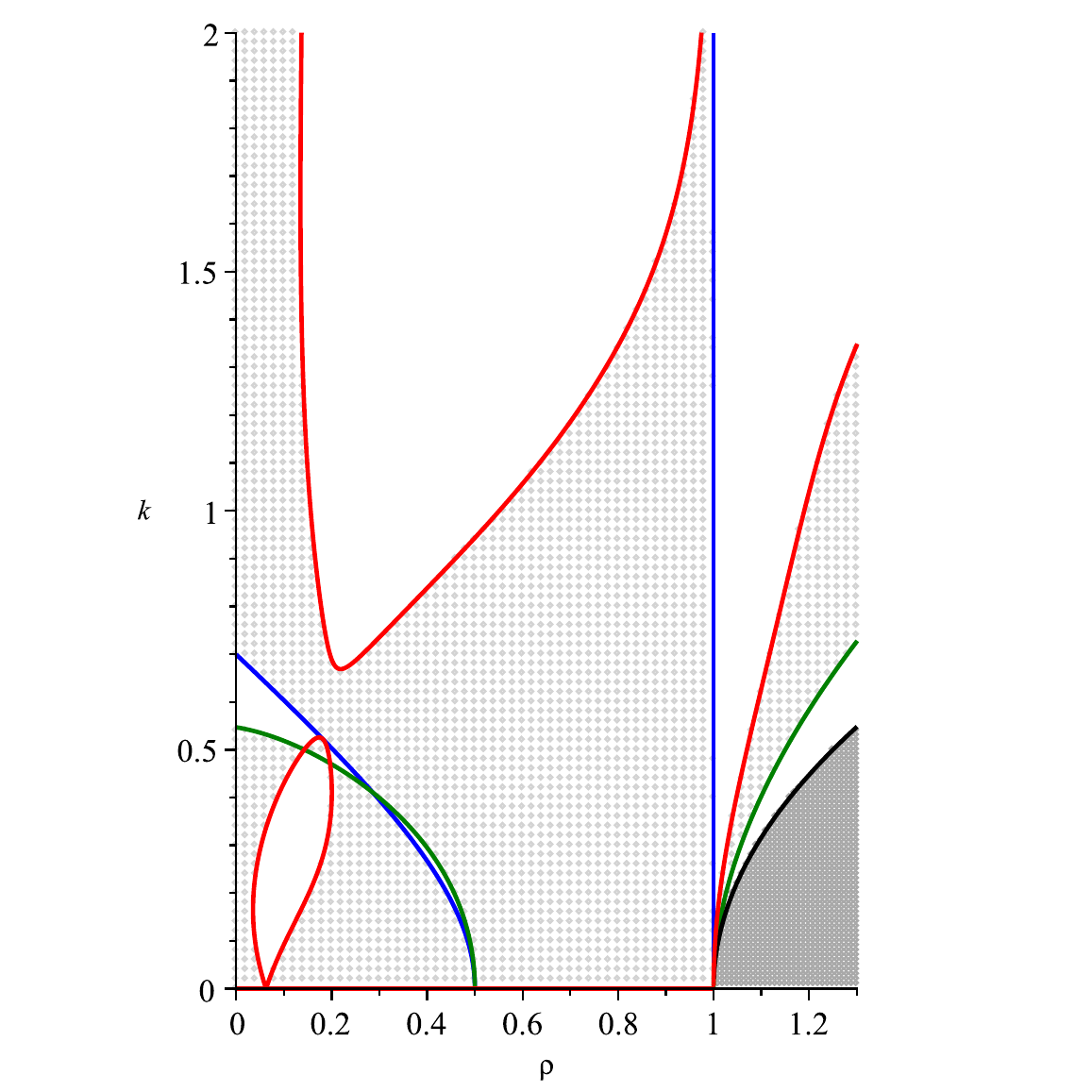}
  \caption{$h_1\!=\!4,\, h_2\!=\!1$}
  \label{fig:Fig1m}
\end{subfigure}\hspace{-6ex}
\begin{subfigure}[b]{0.29\textwidth}
  \includegraphics[width=\linewidth]{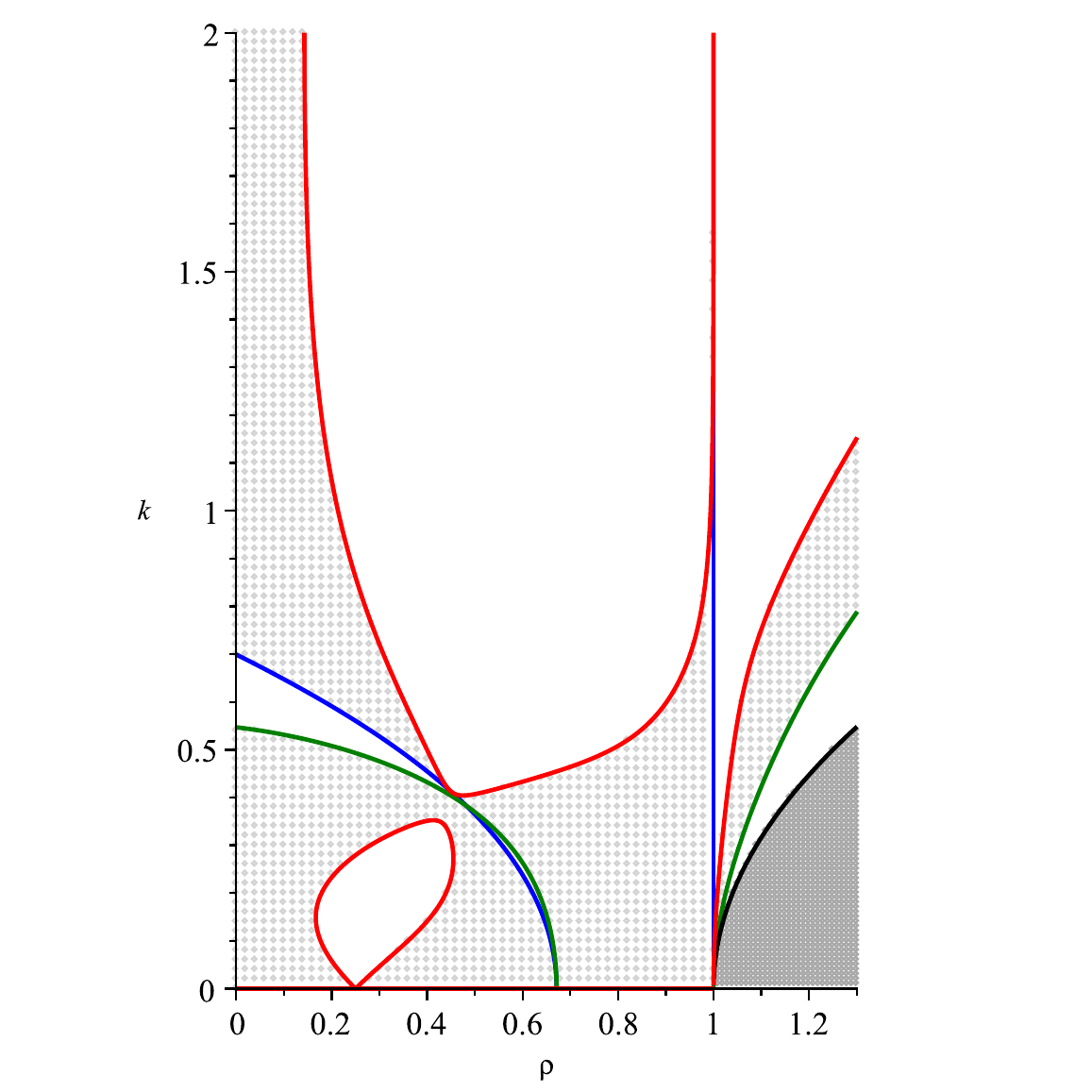}
  \caption{$h_1\!=\!4,\, h_2\!=\!2$}
  \label{fig:Fig1n}
\end{subfigure}\hspace{-6ex}
\begin{subfigure}[b]{0.29\textwidth}
  \includegraphics[width=\linewidth]{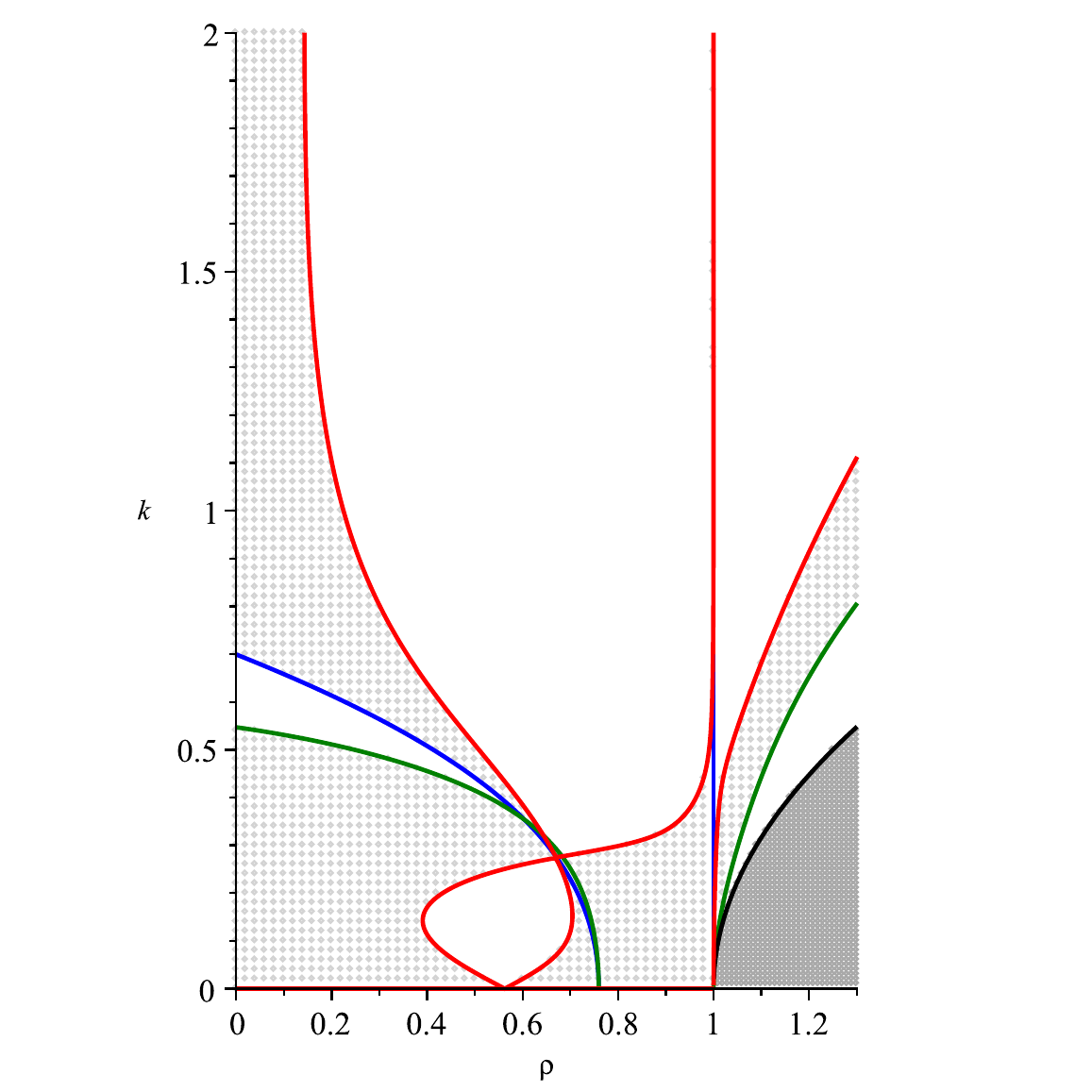}
  \caption{$h_1\!=\!4,\, h_2\!=\!3$}
  \label{fig:Fig1o}
\end{subfigure}\hspace{-6ex}
\begin{subfigure}[b]{0.29\textwidth}
  \includegraphics[width=\linewidth]{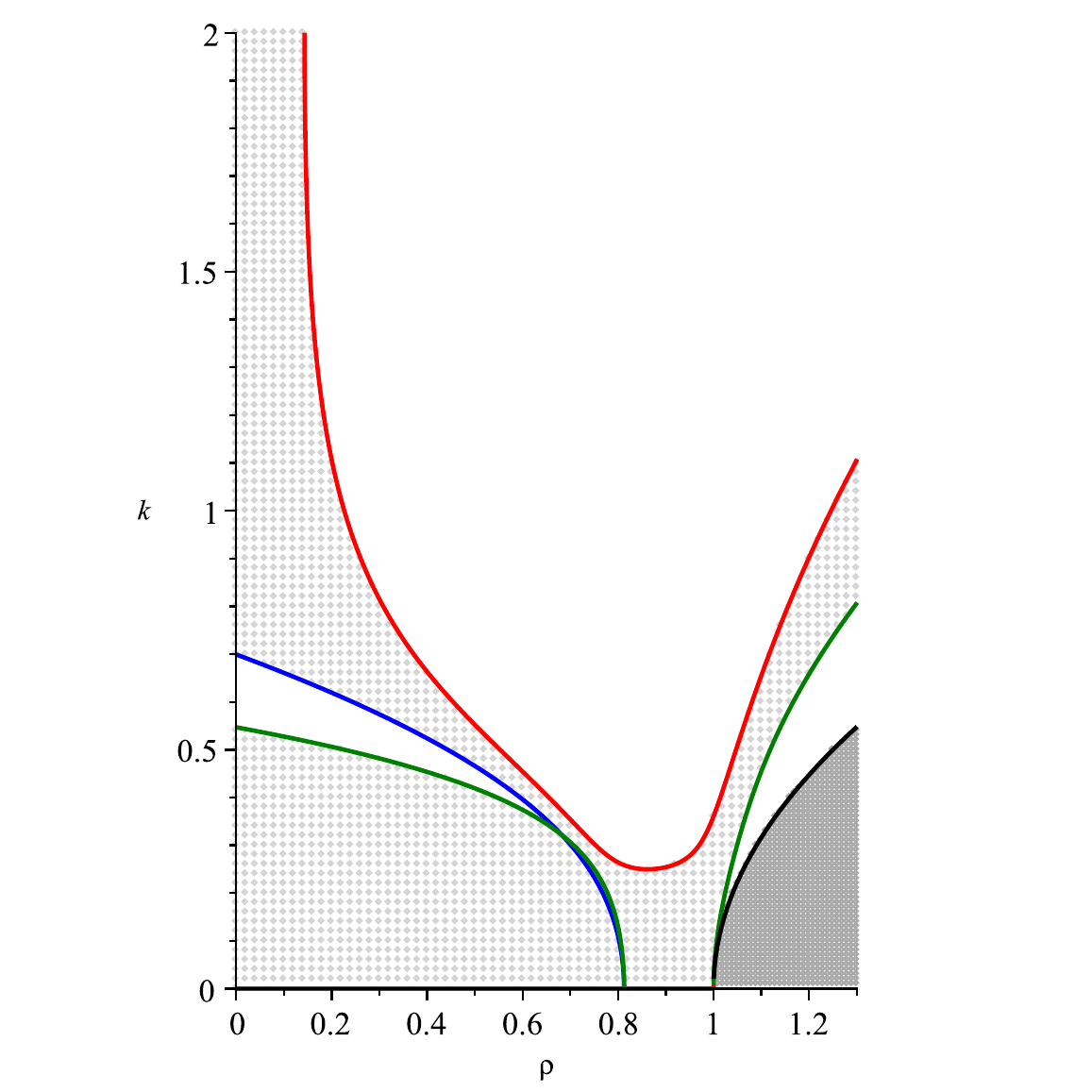}
  \caption{$h_1\!=\!4,\, h_2\!=\!4$}
  \label{fig:Fig1p}
\end{subfigure}

\caption{Modulational stability diagrams for all combinations of $h_{1}, h_{2} \in \{1, 2, 3, 4\}$.}
\label{fig:Fig1}
\end{figure}


\begin{figure}
\centering

\hspace{-5ex}
\begin{subfigure}[b]{0.29\textwidth}
  \includegraphics[width=\linewidth]{Fig_1a.pdf}
  \caption{$h_1\!=\!1,\, h_2\!=\!1$}
  \label{fig:Fig2a}
\end{subfigure}\hspace{-6ex}
\begin{subfigure}[b]{0.29\textwidth}
  \includegraphics[width=\linewidth]{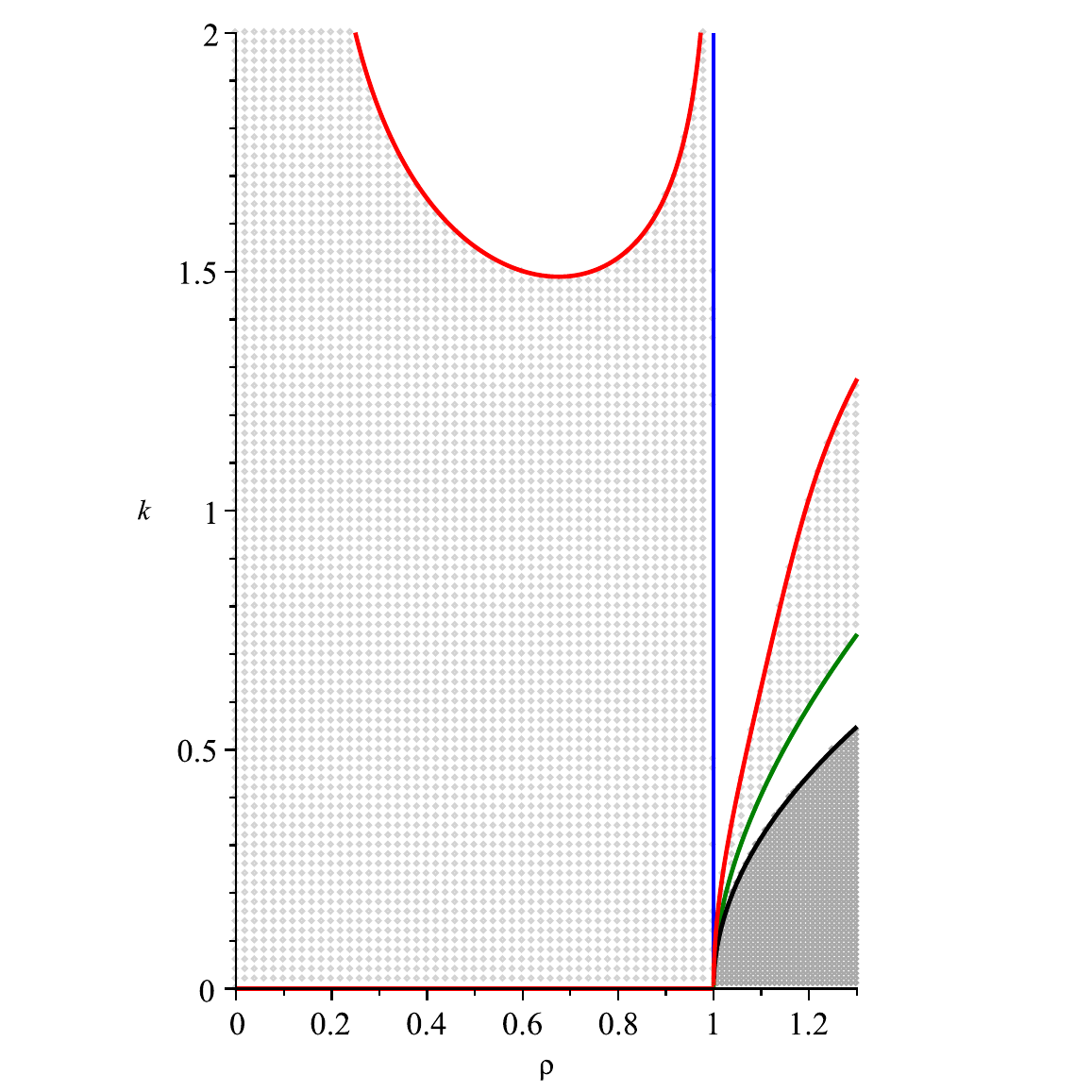}
  \caption{$h_1\!=\!1,\, h_2\!=\!5$}
  \label{fig:Fig2b}
\end{subfigure}\hspace{-6ex}
\begin{subfigure}[b]{0.29\textwidth}
  \includegraphics[width=\linewidth]{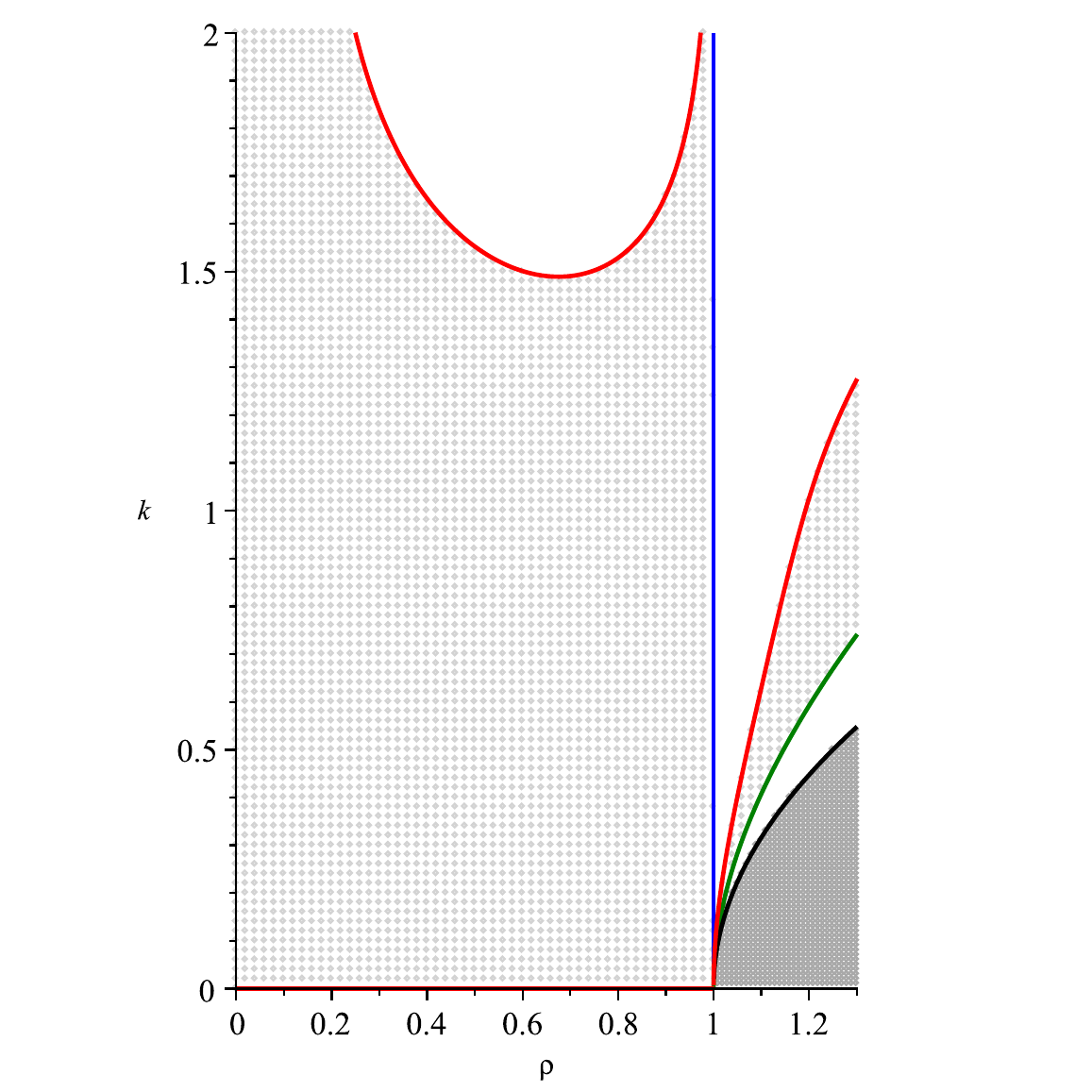}
  \caption{$h_1\!=\!1,\, h_2\!=\!9$}
  \label{fig:Fig2c}
\end{subfigure}\hspace{-6ex}
\begin{subfigure}[b]{0.29\textwidth}
  \includegraphics[width=\linewidth]{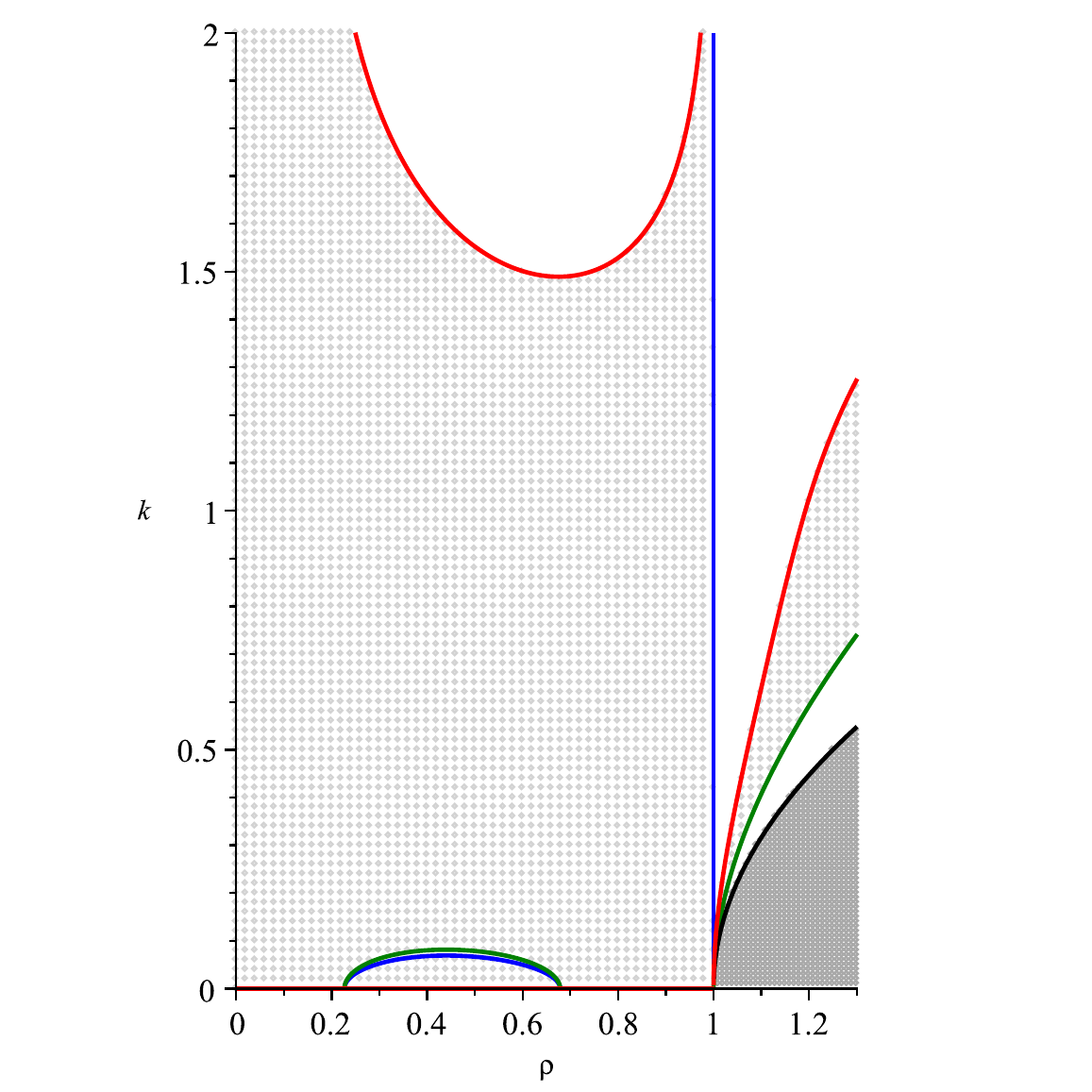}
  \caption{$h_1\!=\!1,\, h_2\!=\!13$}
  \label{fig:Fig2d}
\end{subfigure}

\vspace{0.5em}

\hspace{-5ex}
\begin{subfigure}[b]{0.29\textwidth}
  \includegraphics[width=\linewidth]{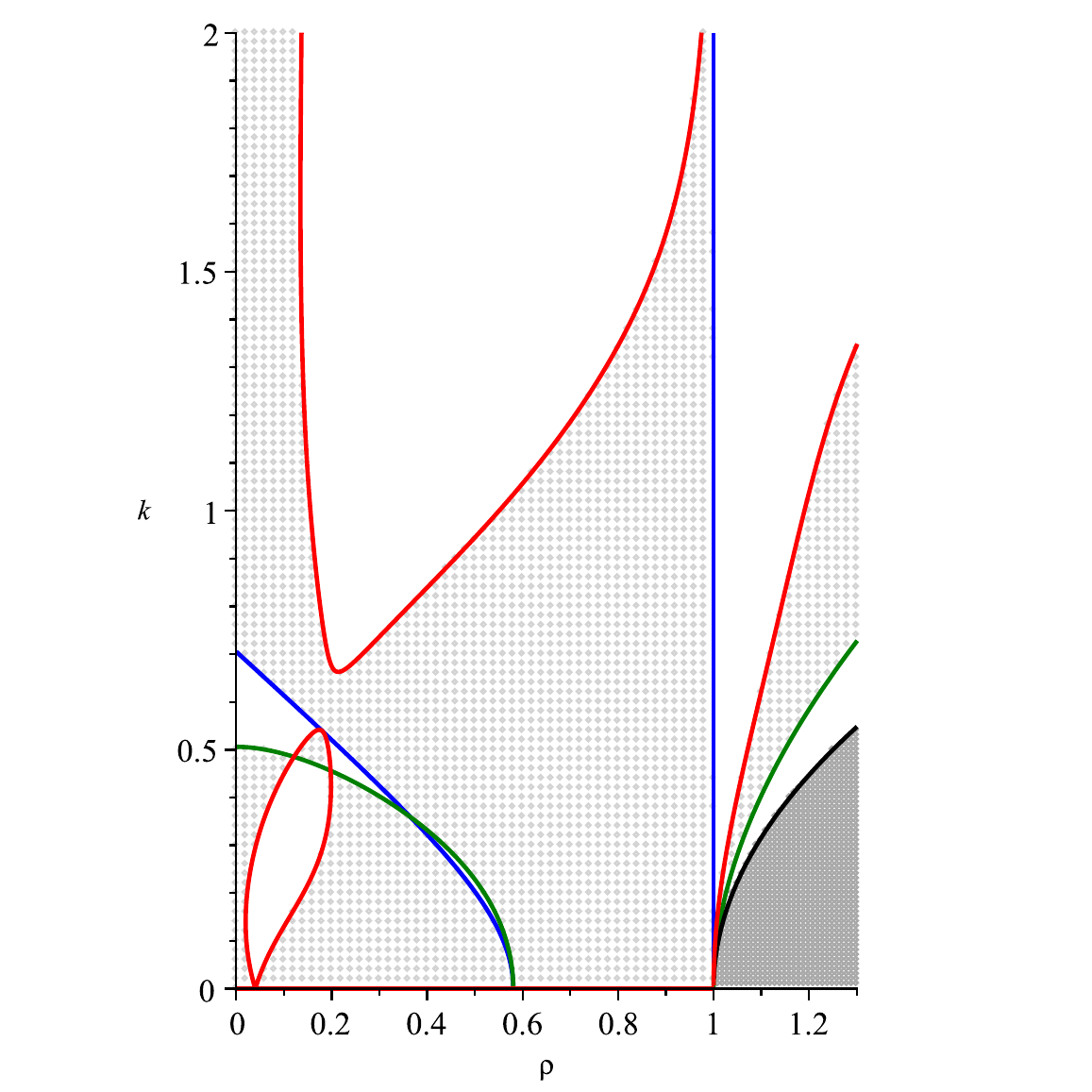}
  \caption{$h_1\!=\!5,\, h_2\!=\!1$}
  \label{fig:Fig2e}
\end{subfigure}\hspace{-6ex}
\begin{subfigure}[b]{0.29\textwidth}
  \includegraphics[width=\linewidth]{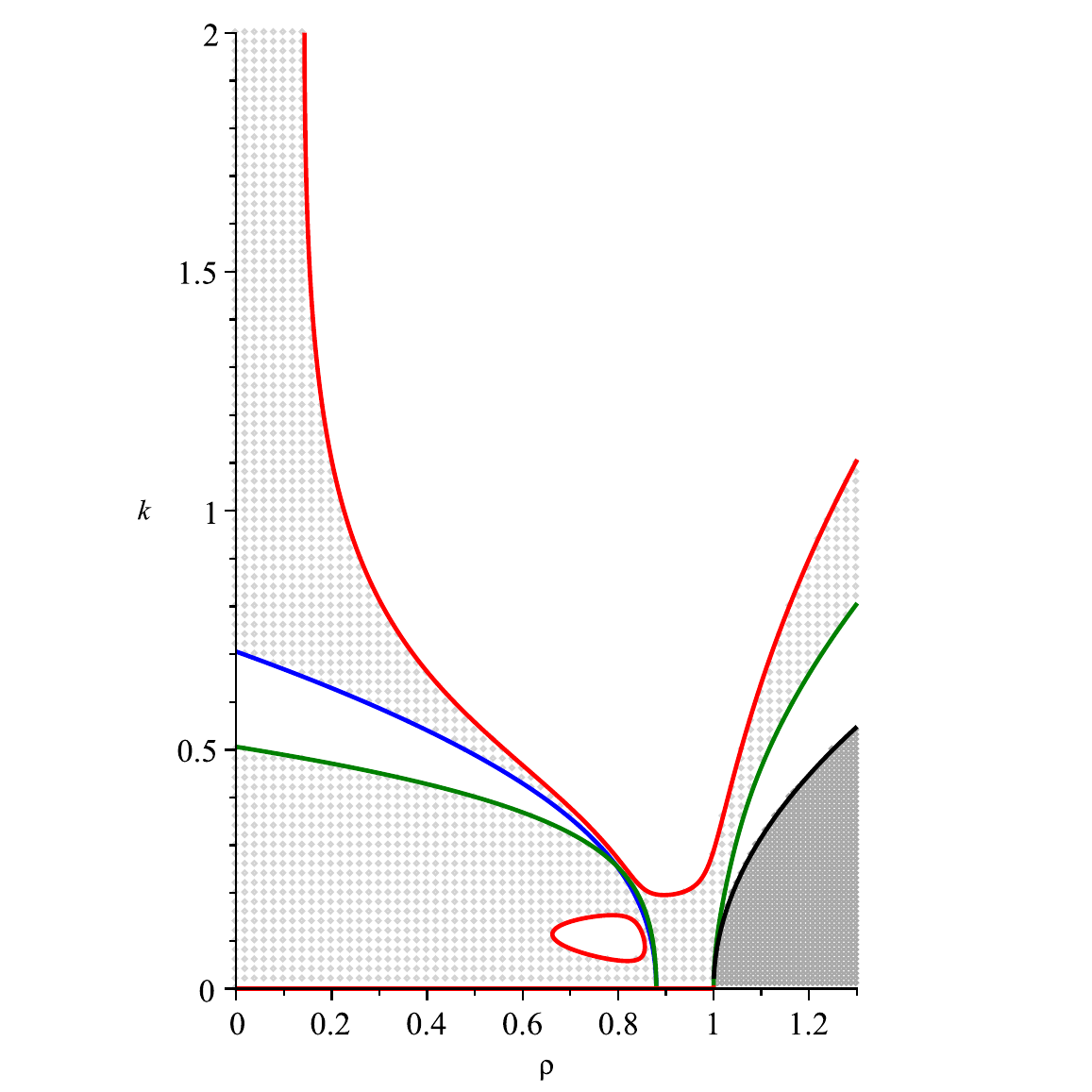}
  \caption{$h_1\!=\!5,\, h_2\!=\!5$}
  \label{fig:Fig2f}
\end{subfigure}\hspace{-6ex}
\begin{subfigure}[b]{0.29\textwidth}
  \includegraphics[width=\linewidth]{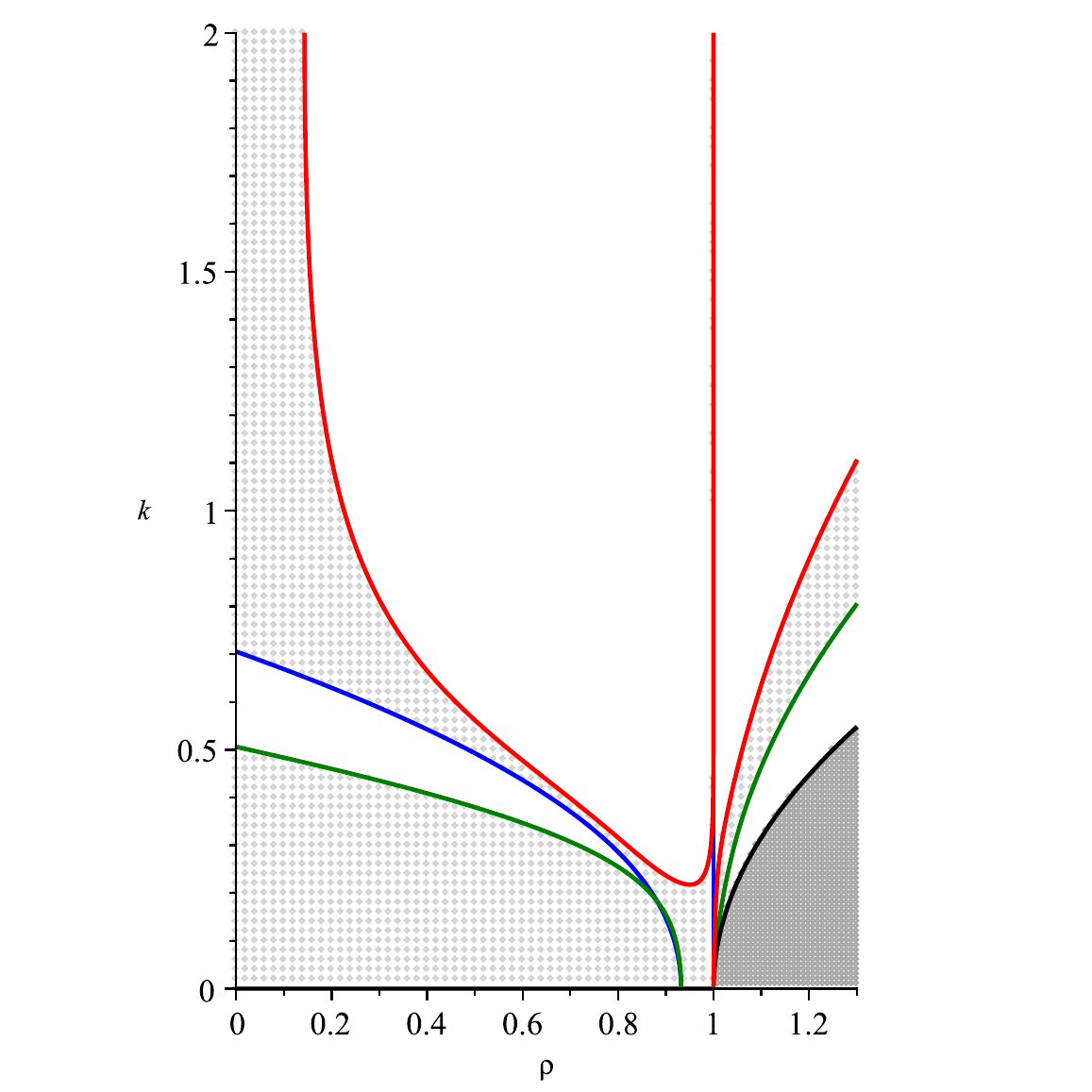}
  \caption{$h_1\!=\!5,\, h_2\!=\!9$}
  \label{fig:Fig2g}
\end{subfigure}\hspace{-6ex}
\begin{subfigure}[b]{0.29\textwidth}
  \includegraphics[width=\linewidth]{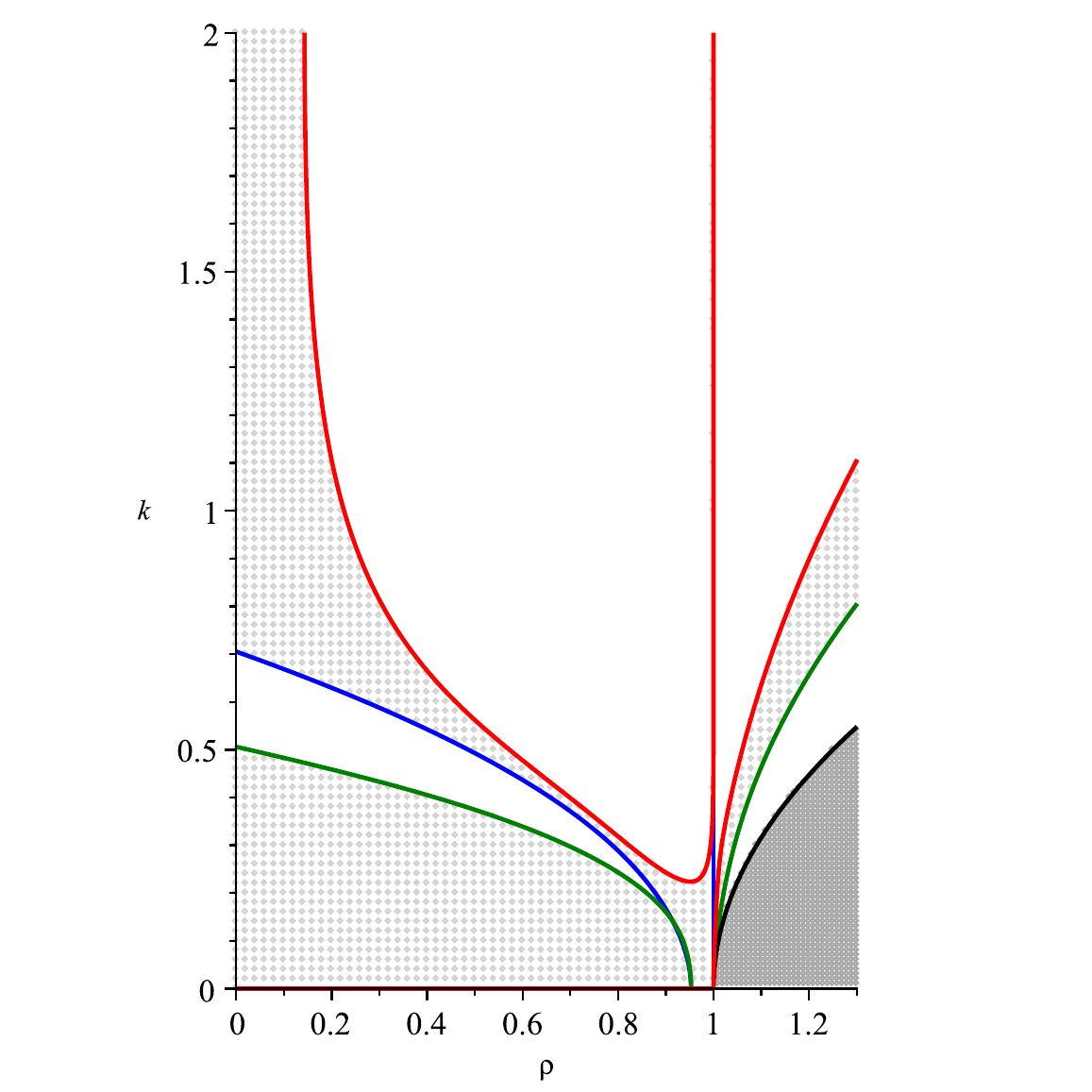}
  \caption{$h_1\!=\!5,\, h_2\!=\!13$}
  \label{fig:Fig2h}
\end{subfigure}

\vspace{0.5em}

\hspace{-5ex}
\begin{subfigure}[b]{0.29\textwidth}
  \includegraphics[width=\linewidth]{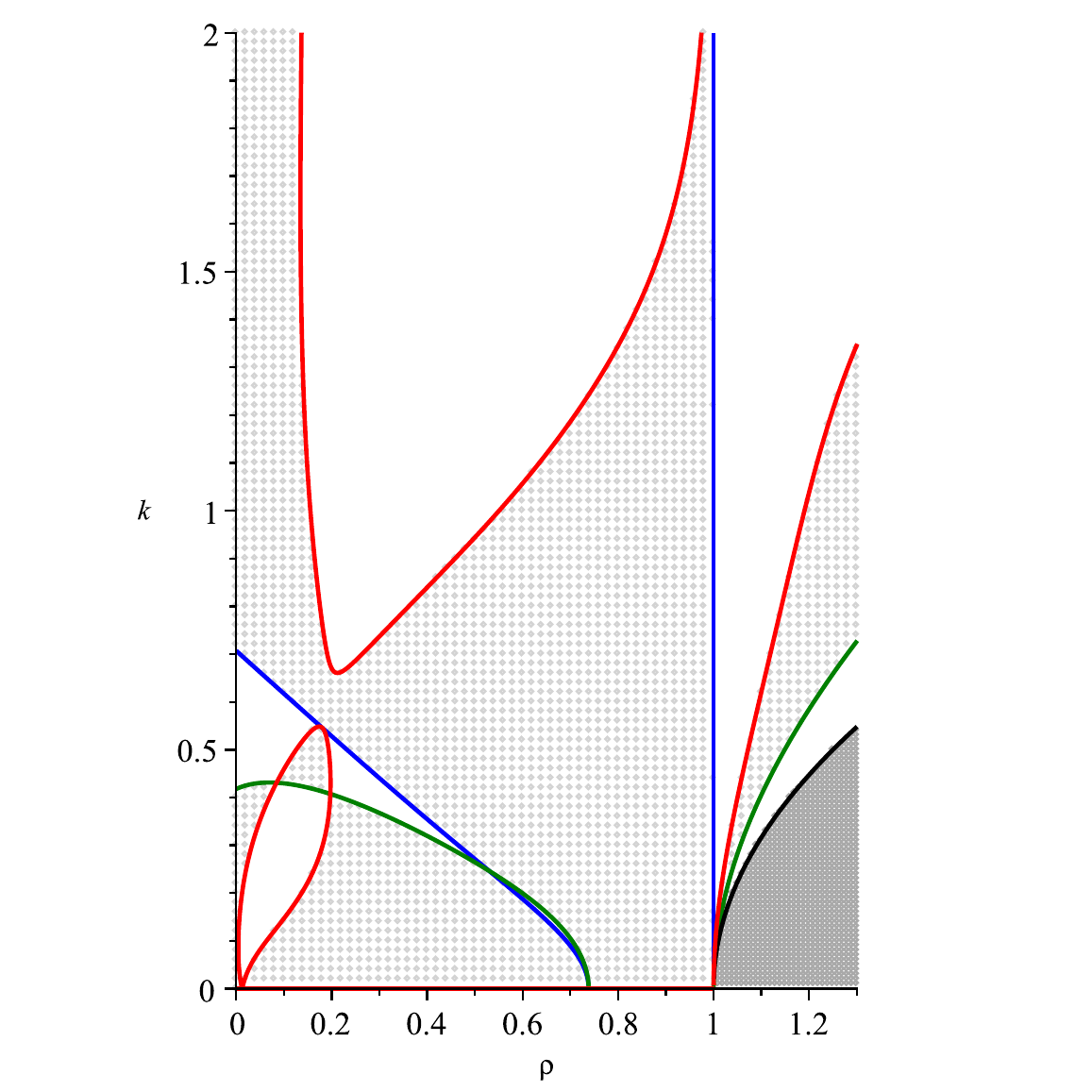}
  \caption{$h_1\!=\!9,\, h_2\!=\!1$}
  \label{fig:Fig2i}
\end{subfigure}\hspace{-6ex}
\begin{subfigure}[b]{0.29\textwidth}
  \includegraphics[width=\linewidth]{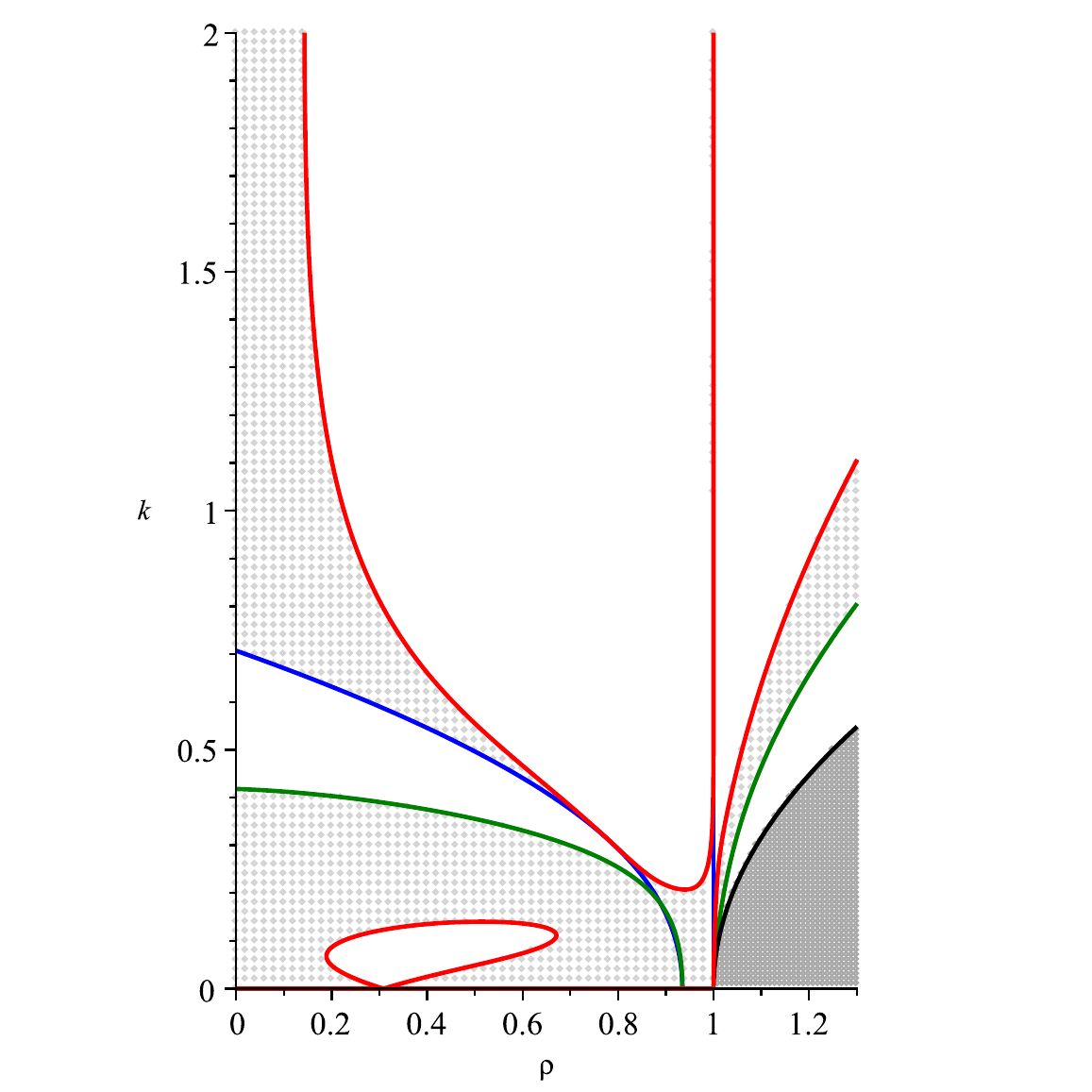}
  \caption{$h_1\!=\!9,\, h_2\!=\!5$}
  \label{fig:Fig2j}
\end{subfigure}\hspace{-6ex}
\begin{subfigure}[b]{0.29\textwidth}
  \includegraphics[width=\linewidth]{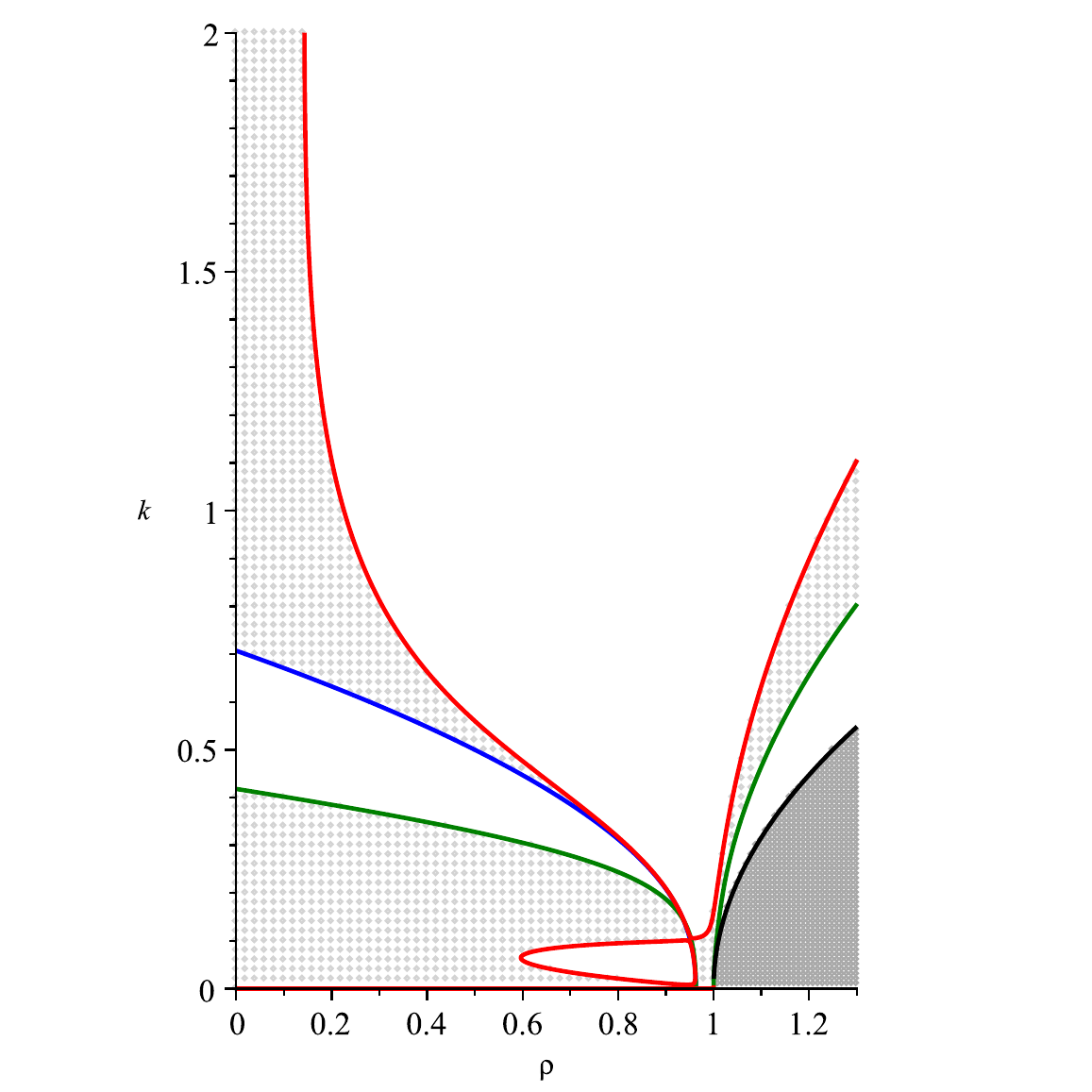}
  \caption{$h_1\!=\!9,\, h_2\!=\!9$}
  \label{fig:Fig2k}
\end{subfigure}\hspace{-6ex}
\begin{subfigure}[b]{0.29\textwidth}
  \includegraphics[width=\linewidth]{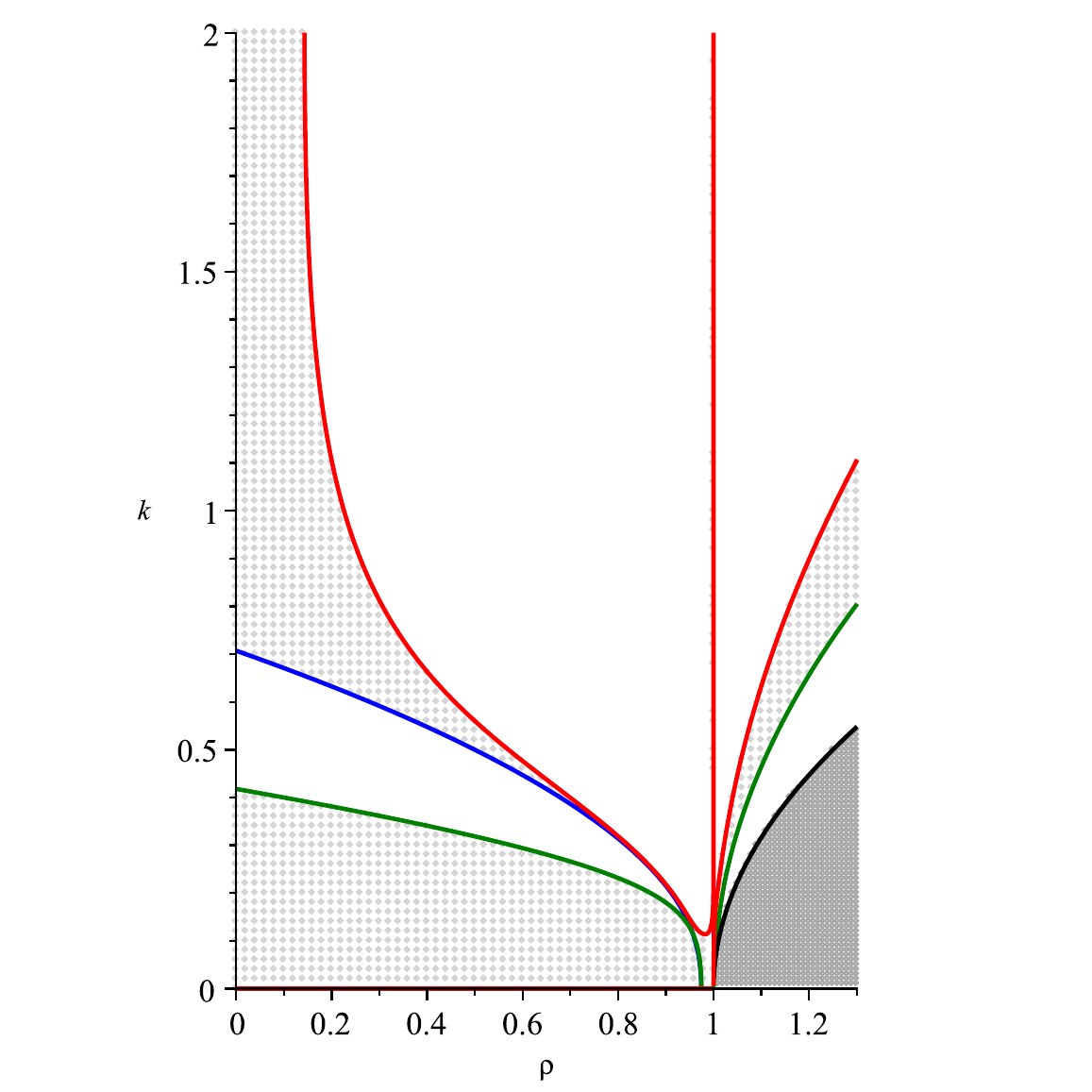}
  \caption{$h_1\!=\!9,\, h_2\!=\!13$}
  \label{fig:Fig2l}
\end{subfigure}

\vspace{0.5em}

\hspace{-5ex}
\begin{subfigure}[b]{0.29\textwidth}
  \includegraphics[width=\linewidth]{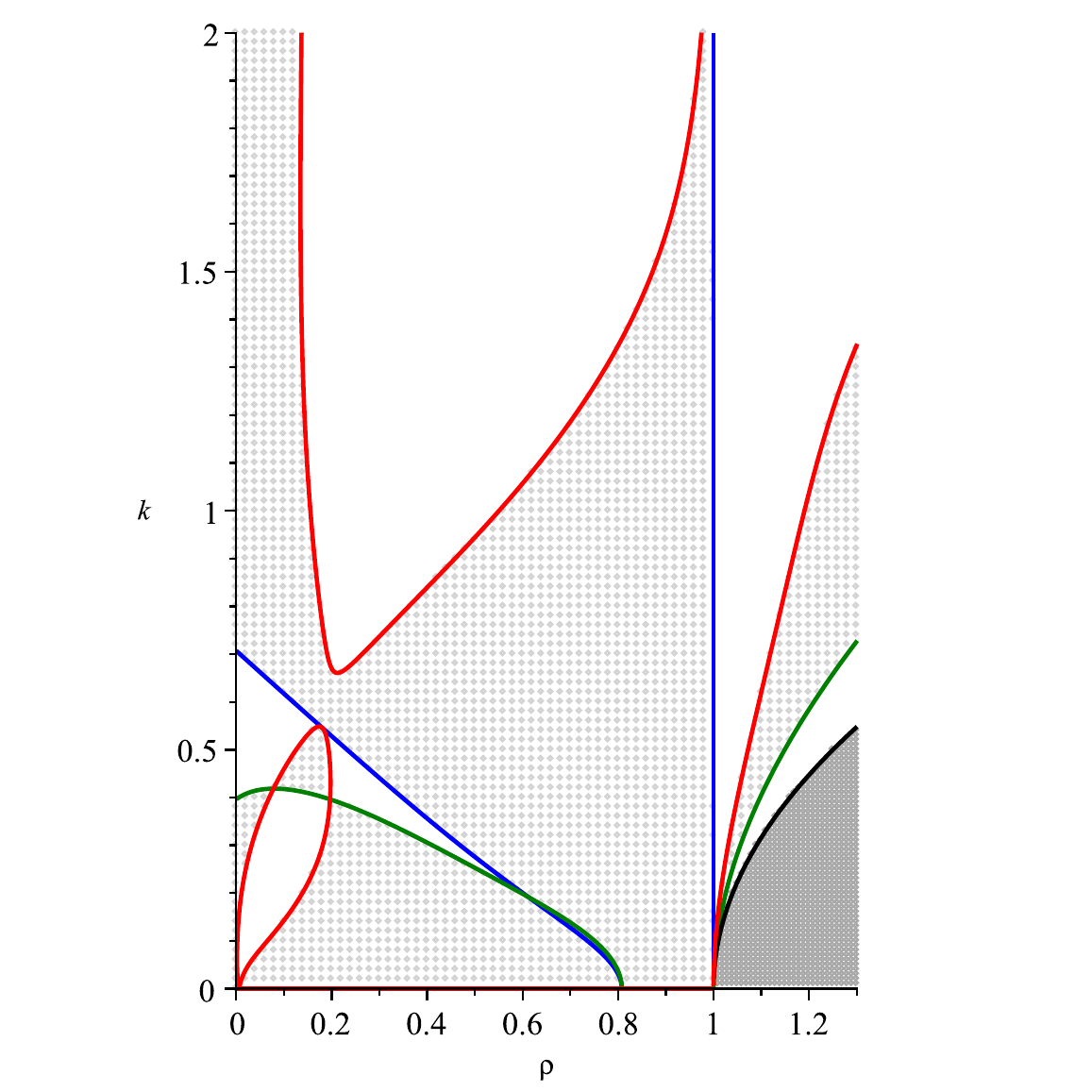}
  \caption{$h_1\!=\!13,\, h_2\!=\!1$}
  \label{fig:Fig2m}
\end{subfigure}\hspace{-6ex}
\begin{subfigure}[b]{0.29\textwidth}
  \includegraphics[width=\linewidth]{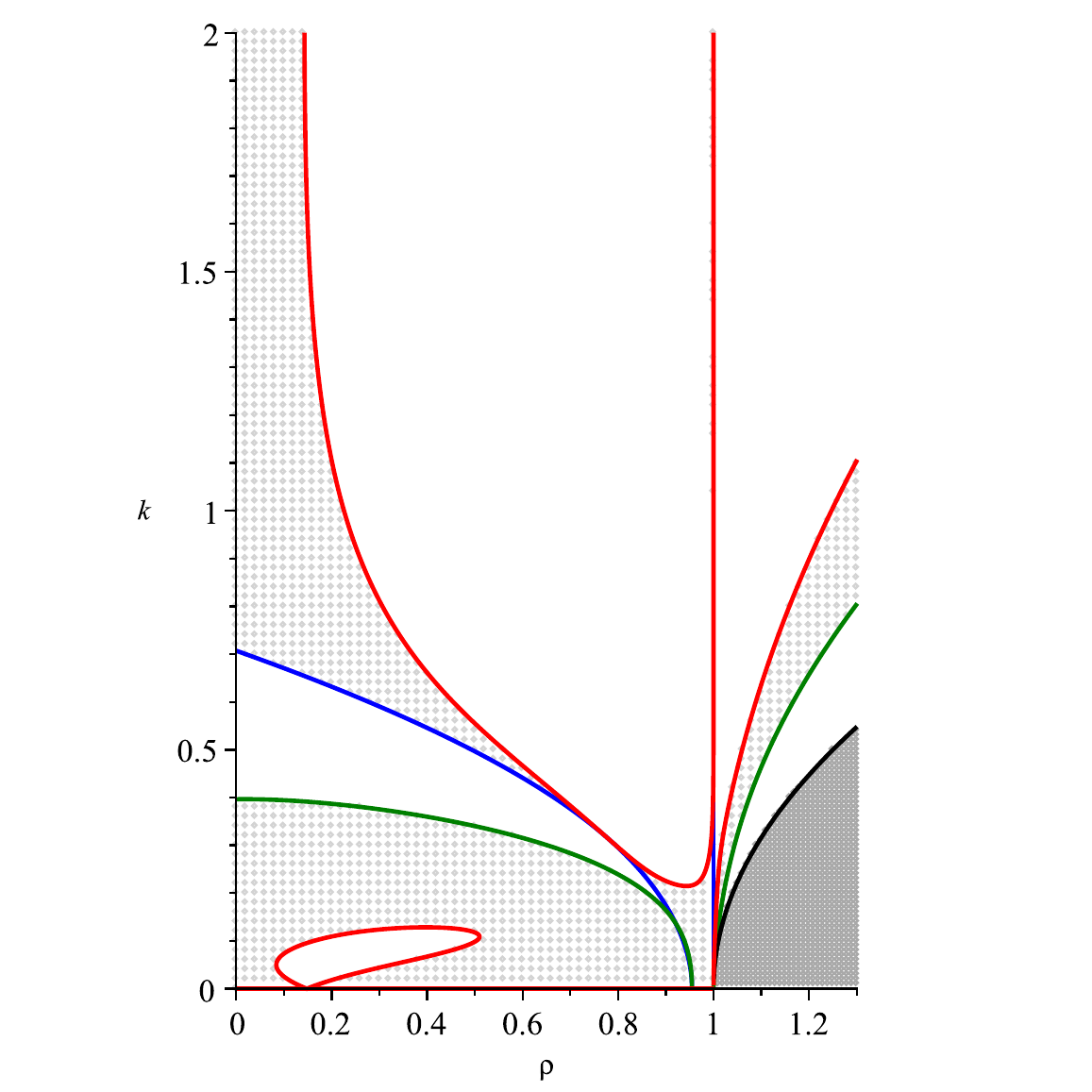}
  \caption{$h_1\!=\!13,\, h_2\!=\!5$}
  \label{fig:Fig2n}
\end{subfigure}\hspace{-6ex}
\begin{subfigure}[b]{0.29\textwidth}
  \includegraphics[width=\linewidth]{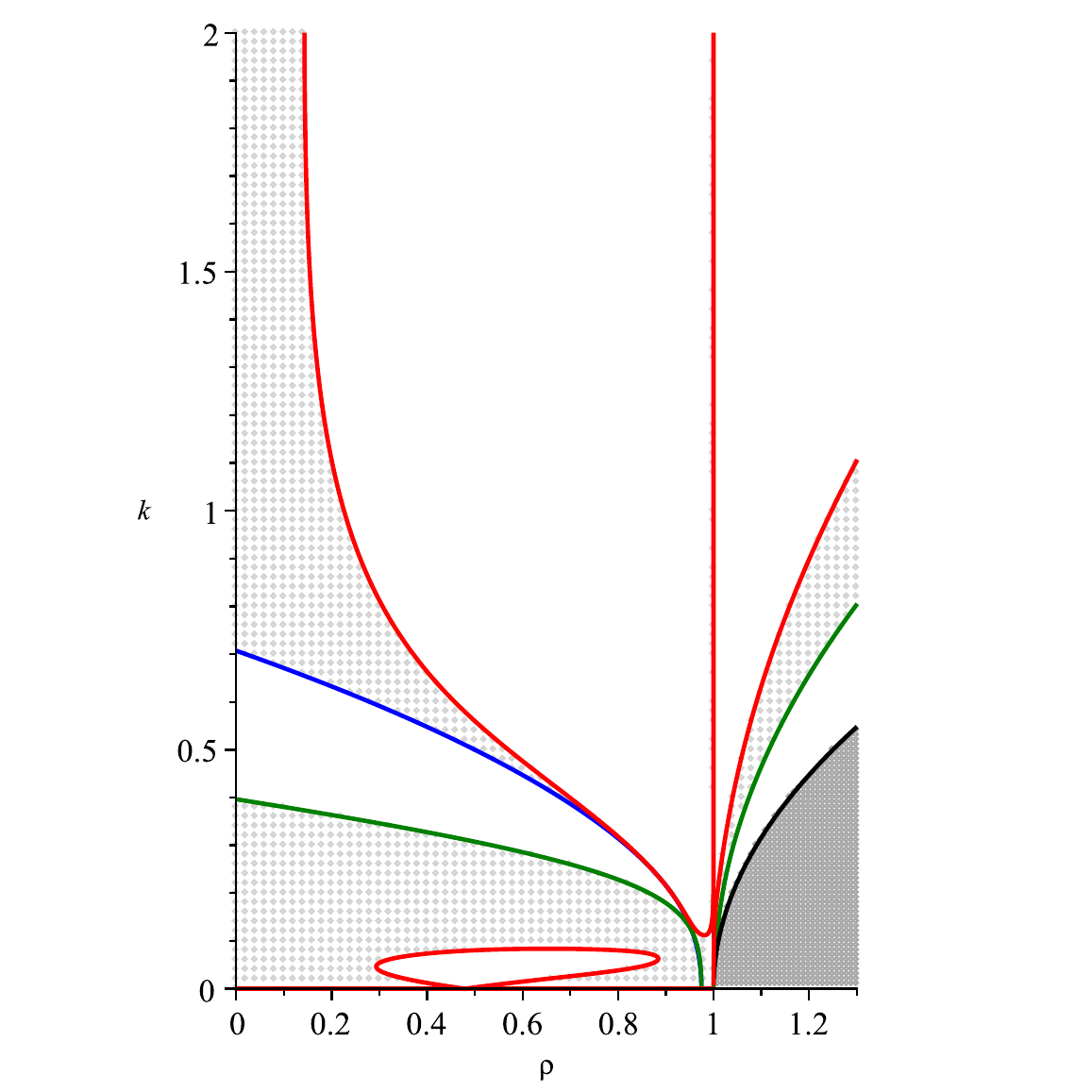}
  \caption{$h_1\!=\!13,\, h_2\!=\!9$}
  \label{fig:Fig2o}
\end{subfigure}\hspace{-6ex}
\begin{subfigure}[b]{0.29\textwidth}
  \includegraphics[width=\linewidth]{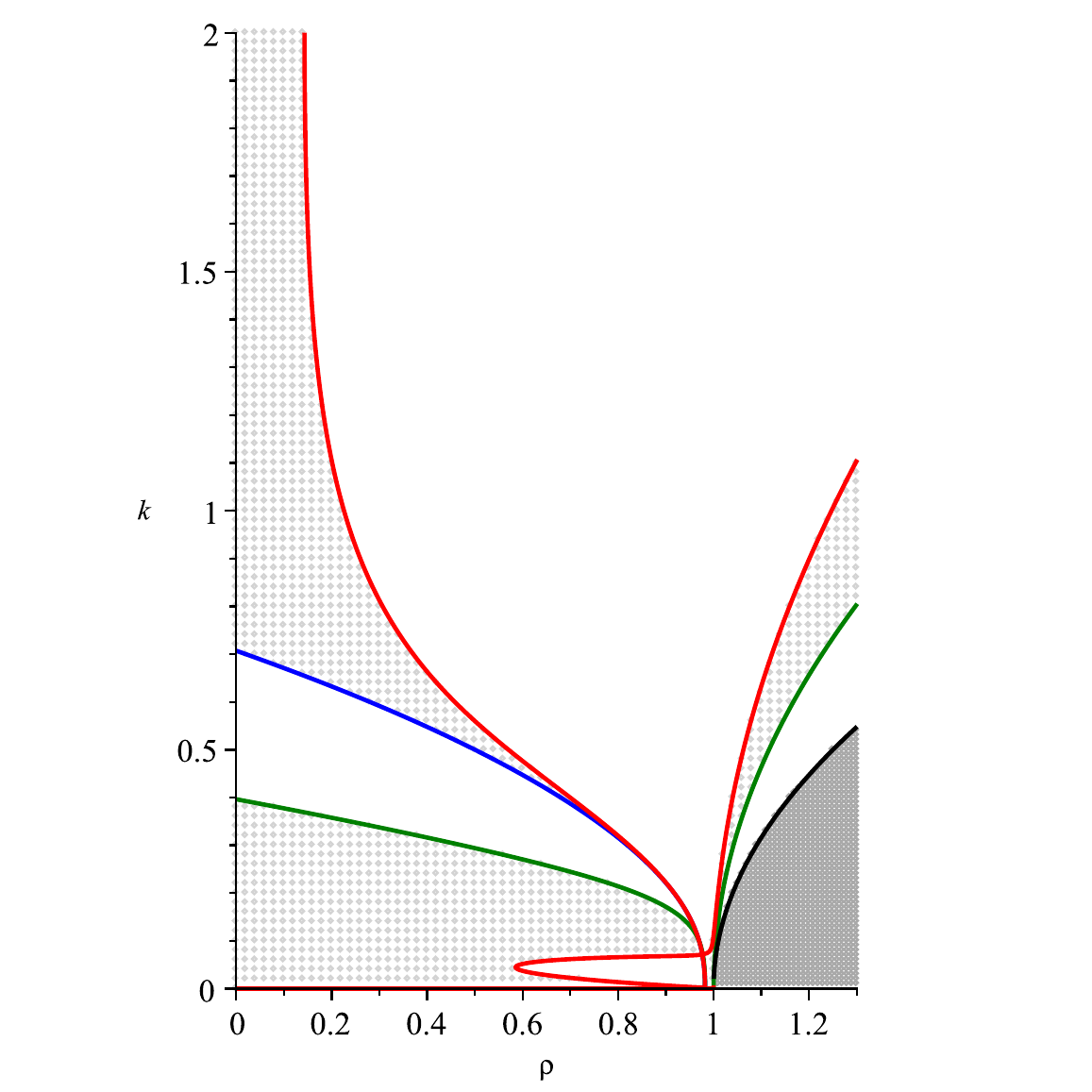}
  \caption{$h_1\!=\!13,\, h_2\!=\!13$}
  \label{fig:Fig2p}
\end{subfigure}

\caption{Modulational stability diagrams for all combinations of $h_{1}, h_{2} \in \{1, 5, 9, 13\}$.}
\label{fig:Fig2}
\end{figure}

Figures~\ref{fig:Fig1} and \ref{fig:Fig2} present $4\times 4$ matrix-format stability diagrams for various combinations of lower-layer thickness $h_1$ and upper-layer thickness $h_2$.
Figure~\ref{fig:Fig1} covers $h_1,h_2\in\{1,2,3,4\}$, allowing one to trace topological changes with small increments in the system geometry.
Figure~\ref{fig:Fig2} covers $h_1,h_2\in\{1,5,9,13\}$, enabling the study of cases with pronounced thickness contrasts and analysis of limiting regimes.
The case $h_1=1, h_2=1$ appears in both matrices (Figs.~\ref{fig:Fig1}a and \ref{fig:Fig2}a), facilitating the understanding of stability/instability evolution with changing geometric parameters.
These diagrams are analysed in detail in the following sections.

Examining the diagonal cells of the matrices in Figs.~\ref{fig:Fig1} and \ref{fig:Fig2} reveals the evolution of characteristic curves and the stability/instability ratio as $h_1$ and $h_2$ increase synchronously.
This approach identifies key topological transitions in a symmetric two-layer system as it approaches the HS--HS limit (see Sec.~\ref{sec:h1=h2}).
Analysis of the first column tracks the effect of increasing $h_1$ at fixed $h_2=1$, characterising transitions in an asymmetric system towards the HS--La limit (see Sec.~\ref{sec:h2=constant}).
Analysis of the first row shows the changes occurring when $h_2$ increases at fixed $h_1=1$, identifying transitions towards the La--HS limit (see Sec.~\ref{sec:h1=constant}).
In this context, several recurrent topological structures will be encountered, namely the \emph{upper} stability region, the \emph{stability corridor}, the \emph{loop}, and the \emph{cut}, which will serve as reference features in the subsequent analysis.

\subsection{Modulational stability diagrams for equal layer thicknesses}
\label{sec:h1=h2}

We now examine how the topology of modulational stability diagrams changes as the equal layer thicknesses $h_1=h_2=h$ increase.
The diagonal entries of the matrix plots in Figs.~\ref{fig:Fig1} and \ref{fig:Fig2} trace this evolution from thin layers to the deep-water limit (HS--HS).
In all cases, the principal curves are: $J=0$ (red), $\omega''=0$ (green), $\omega=0$ (black), and $J\to\infty$ (blue).
The sign combinations of $J$ and $\omega''$ determine stability: instability of the Benjamin--Feir type for $J>0$, $\omega''>0$; the opposite instability for $J<0$, $\omega''<0$; and stability when the two effects compensate ($J>0$, $\omega''<0$ or $J<0$, $\omega''>0$).
In the following analysis we will also refer to recurrent localized features emerging from these curves, namely the \emph{upper} stability region, the \emph{stability corridor}, and the \emph{loop}.

\medskip
\noindent\textbf{Small thicknesses ($h=1,2$).}
For $h=1$ (Fig.~\ref{fig:Fig1}a), the diagram exhibits the basic topology: a U-shaped $J=0$ curve, a green $\omega''=0$ line starting at $(\rho=1,k=0)$, and a black $\omega=0$ curve slightly below it.
The region below $J=0$ and bounded on the right by $\omega''=0$ is modulationally unstable (Benjamin--Feir type), while the upper region above $J=0$ is stable.
A second stability region occurs between $\omega''=0$ and $\omega=0$ for $\rho>1$.

At $h=2$ (Fig.~\ref{fig:Fig1}f), all four characteristic curves appear, with the $J\to\infty$ line and a new branch of $\omega''=0$ forming, at small $\rho$, a narrow stability corridor separating two instability zones.
The overall arrangement to the right of $\rho=1$ remains close to the $h=1$ case.

\medskip
\noindent\textbf{Intermediate thicknesses ($h=3,4,5$).}
For $h=3$ and $h=4$ (Figs.~\ref{fig:Fig1}k,p), the blue and green curves diverge further at small $\rho$, enlarging the range of stability by compensation.
For $h=3$, their intersections enclose a lens-shaped stability region; at $h=4$, the stability corridor shifts rightwards and partially merges with regions to the right of $\rho=1$.
The U-shaped $J=0$ curve deforms and shifts to smaller $k$, lowering and displacing the upper region.

At $h=5$ (Fig.~\ref{fig:Fig2}f), a qualitative change occurs: an isolated closed $J=0$ curve (a loop) appears in the lower-left diagram, enclosing a local stability island inside an instability zone (isola-bifurcation between $h=4$ and $h=5$).
The stability corridor continues to widen and link distinct stable zones.

\medskip
\noindent\textbf{Large thicknesses and deep-water limit ($h=9,13,\infty$).}
For $h=9$ (Fig.~\ref{fig:Fig2}k), the stability corridor closes into an elongated loop bounded by $J\to\infty$ and $\omega''=0$, forming a stability island in $(\rho,k)$ space.
The $J=0$ loop persists inside the main instability zone.
At $h=13$ (Fig.~\ref{fig:Fig2}p), the left-side curves $J=0$, $J\to\infty$, and $\omega''=0$ nearly coincide, producing extremely narrow alternating stability/instability bands and stretching the loop towards the $k$-axis.

In the limit $h\to\infty$, the configuration converges to the HS--HS topology \cite{Nayfeh1976}: the loop disappears, $J=0$, $J\to\infty$, and $\omega''=0$ take their deep-water forms, and the $(\rho,k)$ plane splits into large continuous stable and unstable regions without isolated resonance islands.

\subsection{Modulational stability diagrams for varying lower-layer thickness at fixed upper-layer thickness}
\label{sec:h2=constant}

We now investigate the influence of increasing lower-layer thickness $h_1$ while keeping the upper-layer thickness fixed at $h_2 = 1$.
The first columns of the matrix plots in Figs.~\ref{fig:Fig1} and \ref{fig:Fig2} show the evolution of characteristic curves ($J=0$, $J\to\infty$, $\omega''=0$, $\omega=0$) and the associated stability/instability regions.
As in the symmetric case, stability is determined by the sign combinations of $J$ and $\omega''$, and localised structures such as the \emph{stability corridor}, \emph{loop}, and \emph{upper} stability region are defined in Sec.~\ref{sec:notation} and referred to here without repetition.
A distinctive feature of all asymmetric configurations is that $\rho = 1$ acts as a universal boundary between stability and instability, since
\(
\lim_{\rho \to 1} J = -\infty
\)
for $h_{1} \neq h_{2}$, regardless of $k$, as follows from Eq.~\eqref{J_lim}.

\medskip
\noindent\textbf{Small lower-layer thicknesses ($h_1=2,3$).}
Transitioning from the symmetric $h_1=h_2=1$ case to $h_1=2$ (Fig.~\ref{fig:Fig1}e) produces a narrow stability corridor at small $\rho$, bounded by $J\to\infty$ and a new branch of $\omega''=0$.
Compared to the symmetric configuration, the corridor is much narrower due to the limited range of compensation between nonlinearity and dispersion in the asymmetric geometry.
At $h_1=3$ (Fig.~\ref{fig:Fig1}i), a loop appears below the corridor, tangent to $k=0$ and surrounded by an instability zone.
The corridor itself develops two intersections between $J\to\infty$ and $\omega''=0$, segmenting it into alternating stable and unstable strips.
The upper stability region shifts and elongates towards the loop, suggesting possible merging for larger $h_1$.

\medskip
\noindent\textbf{Intermediate lower-layer thicknesses ($h_1=4,5$).}
For $h_1=4$ (Fig.~\ref{fig:Fig1}m), the corridor widens as $J\to\infty$ and $\omega''=0$ diverge, extending the parameter range for stability by resonance compensation.
The loop enlarges, tilts, and shifts rightwards along $k=0$, indicating stability for higher $\rho$ and longer waves.
Partial overlap between the loop and corridor creates a composite stability domain with internal segmentation.
At $h_1=5$ (Fig.~\ref{fig:Fig2}e), these tendencies strengthen: the loop expands further and its contact point with $k=0$ moves right; the corridor stretches towards larger $\rho$ and overlaps more with the loop, forming an extended multi-mechanism stability region separated from the upper zone by the $J\to\infty$ resonance line.

\medskip
\noindent\textbf{Large lower-layer thicknesses ($h_1=9,13$).}
For $h_1=9$ (Fig.~\ref{fig:Fig2}i), the loop continues to grow and retain its tilt, with the contact point shifting further along $k=0$.
The corridor extends towards higher $\rho$, increasing its overlap with the loop and producing a complex composite stability structure with narrow intervening instability bands.
At $h_1=13$ (Fig.~\ref{fig:Fig2}m), the $J=0$, $J\to\infty$, and $\omega''=0$ curves in the left part of the diagram move closer together, narrowing the corridor and increasing sensitivity to parameter variations, while the loop remains prominent.

\medskip
\noindent\textbf{Deep lower-layer limit ($h_1\to\infty$).}
In the limit of an infinitely deep lower layer, the configuration approaches the HS--La model \cite{AvramenkoNarad2025}.
Here, the loop root lies at $(\rho,k)=(0,0)$, the corridor broadens and extends to $(1,0)$, and the upper stability region approaches but does not merge with the loop due to the intervening $J\to\infty$ resonance barrier.
With the influence of the lower boundary removed, stability and instability are controlled solely by the interfacial mode and the upper layer, without additional isolated structures.

\subsection{Modulational stability diagrams for varying upper-layer thickness at fixed lower-layer thickness}
\label{sec:h1=constant}

We now examine how the topology of modulational stability diagrams changes when the upper-layer thickness $h_2$ increases while the lower-layer thickness remains fixed at $h_1=1$.
The first rows of Figs.~\ref{fig:Fig1} and \ref{fig:Fig2} trace this evolution.
The principal curves ($J=0$, $J\to\infty$, $\omega''=0$, $\omega=0$) and the sign combinations of $J$ and $\omega''$ determine the stability regimes, as described in Sec.~\ref{sec:notation}.
In this asymmetric setting, most recurrent structures disappear: only the \emph{upper} stability region persists, intersected by the universal boundary $\rho=1$ corresponding to the vertical $J\to\infty$ resonance line. For sufficiently large $h_2$, a new degenerate structure, the \emph{cut}, emerges from a local degeneracy of $\omega''$ without a change in dispersion type.

\medskip
\noindent\textbf{Small to moderate upper-layer thicknesses ($h_2=2,3,4,5,9$).}
For all these cases (Figs.~\ref{fig:Fig1}b--d, \ref{fig:Fig2}b,c), the U-shaped $J=0$ branch lies in the left part of the diagram, while an additional $J=0$ branch emerges from $(\rho=1,k=0)$ and interacts with the vertical $J\to\infty$ line, producing a stability region absent in symmetric configurations.
The lower instability zone (Benjamin--Feir type) persists, while the U-shaped branch and the upper stability region above it shift slowly towards larger $k$, indicating reduced stability for long waves.
The right-hand stability zone ($J>0$, $\omega''<0$) bounded by $\omega''=0$ and $\omega=0$ changes little with $h_2$.

\medskip
\noindent\textbf{Strongly asymmetric case ($h_2=13$).}
For $h_2=13$ (Fig.~\ref{fig:Fig2}d), a new local structure appears in the lower part of the diagram: a closed $\omega''=0$ contour (cut) enclosing a small modulationally unstable area within the main instability zone.
Unlike standard $\omega''=0$ contours, the cut does not change the sign of $\omega''$; instead, it modifies the shape and connectivity of instability zones and reflects a local degeneracy of the dispersion relation.
This structure is characteristic of strong thickness asymmetry and elongates with increasing $h_2$.
In the limit $h_2\to\infty$, it closes between $(0,0)$ and $(0,1)$, yielding the La--HS configuration \cite{AvramenkoNarad2026}.

\section{Analysis and discussion of the formation and deformation of principal stability regions}

This section reviews the main stability domains in modulational diagrams of a two-layer fluid: a localized \emph{loop}, the \emph{upper} region bounded by $J=0$, and the \emph{corridor} between $J\to\infty$ and $\omega''=0$; in strongly asymmetric cases a degenerate \emph{cut} also appears. Their occurrence, location in the $(\rho,k)$-plane, and underlying physical mechanisms are discussed.

\subsection{Conditions for the appearance and placement of the loop}

The \emph{loop} appears when the lower layer is at least as thick as the upper one and the overall depth is sufficient to approach the deep-water regime for internal waves.
In these cases, the loop originates from solutions of a single condition corresponding to the $J=0$ curve.
This zone represents a localized balance between focusing or defocusing nonlinearity and anomalous or normal dispersion, in which modulational perturbations are effectively neutralized.

For symmetric configurations ($h_1=h_2$), the loop first emerges at relatively large thicknesses (e.g., $h\ge 5$ in our results) and persists for larger $h$.
With strongly asymmetric large-thickness configurations, it can also occur when the lower layer is only slightly thicker than the upper.
In thin-layer cases (e.g., $h_1=2$, $h_2=1$), boundary effects from the free surface and bottom prevent the $J=0$ curve from forming a closed contour, leaving only open structures such as the stability corridor.

The loop typically lies beneath the stability corridor, but in some configurations with a much thicker lower layer it partially overlaps the corridor, producing intricate interleaving of stable and unstable regions.
Its point of contact with the $k=0$ axis—when present—marks a long-wave limit of localized stability.
This contact point shifts with the $h_1/h_2$ ratio: to lower $\rho$ for strongly asymmetric cases, toward $\rho\approx0.5$ for moderately asymmetric ones, and toward $\rho\approx1$ for nearly symmetric layers, without reaching $(0,1)$.
In certain symmetric or near-symmetric large-thickness cases, the loop remains entirely detached from the $k=0$ axis and may approach the corridor boundary instead.

In the limiting HS--HS configuration ($h_1,h_2\to\infty$), the loop vanishes, merging with the $k=0$ axis.
Here, boundary effects are absent and stability properties are determined solely by the global dispersion characteristics, with no isolated localized stability regions.

\subsection{Influence of geometric parameters on the upper stability region}

The \emph{upper} stability region occupies the part of the $(\rho,k)$-plane above the red $J=0$ curve.
For equal layer thicknesses it spans $0.1716 \lesssim \rho \lesssim 5.8275$ as $k \to \infty$, with limits given by
$\rho = \tfrac{2 - \sqrt{2}}{2 + \sqrt{2}}$ and $\rho = \tfrac{2 + \sqrt{2}}{2 - \sqrt{2}}$,
in agreement with the relations in~\eqref{eq:asymptotes}.
For unequal thicknesses the high-$k$ range splits: $0.1716 \leq \rho \leq 1$ contains the main upper region, while $1 \leq \rho \leq 5.8275$ contains a other stability zone.

The shape of the upper region depends asymmetrically on the thickness ratio $h_1/h_2$.
When the lower layer is thicker, its dispersion dominates, producing a downward shift of the lower boundary toward the stability corridor and, in some cases, a ``pulling'' of the upper region toward a coexisting loop (e.g., $h_1>h_2$ with moderate asymmetry).
When the upper layer is thicker, the stabilizing mechanism changes: the loop does not form, the pulling effect disappears, and the geometry of the upper region is altered accordingly.

This asymmetry, seen for various $(h_1,h_2)$ pairs, reflects the different contributions of the two layers to the dispersion of the internal mode.

\subsection{Conditions for emergence, extent, and width of the stability corridor}

The \emph{stability corridor} is defined as the $(\rho,k)$ region bounded by the blue $J=\infty$ and green $\omega''=0$ curves.
It appears only when the lower layer thickness exceeds a certain threshold: for $h_1=1$ the corridor is absent for all tested $h_2$ values, while a very narrow corridor forms at $h_1=2$ and expands with increasing $h_2$.
Further increase to $h_1=3$–$4$ significantly widens the corridor at small $\rho$, extending its reach toward higher $\rho$ values; for large $h_1$ and $h_2$ (e.g., $5$–$13$) it becomes broad and nearly uniform, with intersection points of its boundaries shifting rightward and nearly merging.

When present, the corridor may intersect with a stability loop, producing complex local–global stability patterns.
Its absence at small $h_{1}$ is explained by the dominant influence of the free surface and bottom boundaries, which prevents the formation of a bounded region enclosed by the $J=\infty$ and $\omega''=0$ curves.
As $h_1$ grows, the lower layer increasingly governs the dispersion, leading to the appearance and gradual broadening of the corridor; in the deep-water limit, it attains a wide, homogeneous form with minimal influence from intersection points.

\subsection{The cut as a degenerate stability region}

The \emph{cut} denotes a local topological structure on the modulational stability diagram in the form of a closed branch of the $\omega''=0$ curve, isolating a small region of modulational instability ($J>0$, $\omega''>0$) from the surrounding zone with identical parameter signs.
Unlike standard $\omega''=0$ contours, where $\omega''$ changes sign across the curve, the cut is degenerate: near the point $(k_0,\rho_0)$ with $\omega''=0$, the sign of $\omega''$ remains the same on both sides.
This feature appears for very thick upper layers combined with a thin lower layer, most clearly for $h_1=1$, $h_2=13$ (Fig.~\ref{fig:Fig2}d), where it lies near the bottom of the diagram and touches the $k=0$ axis.
For even thinner lower layers, the effect is expected at smaller $h_2$, potentially yielding more complex degenerate structures.
Such regions are highly sensitive to density variations and may reveal additional mechanisms of modulational stability and instability.

\section*{Conclusions}

A comprehensive topological classification of modulational stability diagrams for wave packets at the interface of a two-layer fluid with interfacial tension has been developed in the $(\rho,k)$ parameter space for a wide range of layer-thickness ratios. The stability landscape is governed by the interplay of four characteristic curves: $J=0$ (change of nonlinearity sign), $\omega''=0$ (change of dispersion type), $J=\infty$ (resonance condition), and $\omega=0$ (linear existence limit). Depending on system geometry, four distinct stability structures are identified: a localized closed loop, a global upper stability region, an elongated corridor between resonance and dispersion curves, and a degenerate cut structure. Each reflects a specific balance between focusing/defocusing nonlinearity and anomalous/normal dispersion.

The loop appears when the lower layer is sufficiently thick, often in symmetric or near-symmetric configurations, and marks a narrow parameter interval of complete nonlinear–dispersive compensation; it vanishes in the deep-water HS–HS limit. The upper stability region changes shape and position with layer-thickness ratio: when the lower layer is thicker, its boundary shifts toward the loop, whereas with a thicker upper layer the loop is absent. The stability corridor forms only above a threshold lower-layer thickness, widens with increasing depths, and can partially overlap with the loop to create multi-component stability zones. The cut arises when a very thick upper layer is combined with a thin lower layer, corresponding to a local degeneracy of $\omega''$ without change in dispersion type.

These results provide a unified interpretation of modulational stability in terms of global and local nonlinear–dispersive balances, identifying conditions where stability is controlled either by localized parameter islands or by global spectral properties. In a companion study (Part~II), we will investigate how the magnitude of interfacial surface tension modifies the topology of stability boundaries, thereby uncovering additional stabilization and destabilization mechanisms that arise from the combined action of capillary and gravitational effects.

\subsection*{Acknowledgments}
Olga Avramenko thanks the Research Council of Lithuania for supporting this work.

\selectlanguage{ukrainian}

\authoru{Ольга Авраменко}{a,b}
\authoru{Володимир Нарадовий}{c}

\begin{center}	
\textbf{НЕСТІЙКІСТЬ БЕНДЖАМІНА–ФЕЙРА МІЖФАЗНИХ ГРАВІТАЦІЙНО-КАПІЛЯРНИХ ХВИЛЬ У ДВОШАРОВІЙ РІДИНІ}
\AuthorPrintu
\affiliation{a}{Національний університет ``Києво-Могилянська академія'', вул. Сковороди, 2, Київ, 04070, Україна}
\affiliation{b}{Університет Вітовта Великого, вул. К.~Донелайчо, 58, Каунас, 44248, Литва}
\affiliation{c}{Центральноукраїнський державний університет імені Володимира Винниченка, вул. Шевченка, 1, Кропивницький, 25006, Україна}
\end{center}
\vskip -0.9em
\begin{abstract}%
У цій роботі подано детальне дослідження модуляційної стійкості хвильових пакетів на межі розділу двошарової ідеальної нестисливої рідини зі скінченною товщиною шарів і міжфазним поверхневим натягом. Аналіз стійкості виконано для широкого діапазону відношень густин та геометричних конфігурацій, що дозволяє побудувати діаграми стійкості на площині $(\rho,k)$, де $\rho$~— відношення густин, а $k$~— хвильове число несної гармоніки. Як критерій стійкості використано індекс Бенджаміна–Фейра, взаємодія якого з кривизною дисперсійного співвідношення визначає момент виникнення модуляційної нестійкості.
Топологія діаграм стійкості виявляє кілька характерних структур: локалізовану замкнену \emph{петлю} стійкості всередині області нестійкості, глобальну \emph{верхню} область стійкості, витягнутий \emph{коридор}, обмежений резонансною та дисперсійною кривими, а також вироджену структуру типу \emph{розріз}, що виникає у сильно асиметричних випадках. Кожна з цих структур відповідає певному фізичному механізму, пов’язаному з балансом між фокусуючою/дефокусуючою нелінійністю та нормальною/аномальною дисперсією.
Систематична зміна товщини шарів дозволяє простежити формування, деформацію та зникнення цих областей, а також їх об’єднання чи сегментацію внаслідок резонансних ефектів. Розглянуто граничні випадки напівнескінченних шарів для узгодження отриманих результатів із відомими конфігураціями, включно з системами типу ``півпростір–шар'', ``шар–півпростір'' та ``півпростір–півпростір''. Особливу увагу приділено впливу симетрії та асиметрії геометрії шарів, що визначає розташування та зв’язність стійких і нестійких областей у параметричному просторі.
Отримані результати формують єдину концептуальну основу для інтерпретації модуляційної стійкості у двошарових рідинах з поверхневим натягом, підкреслюючи як глобальні режими, керовані дисперсією, так і локалізовані острови стійкості. Ця робота становить Частину~I дослідження; Частина~II буде присвячена ролі змінного поверхневого натягу, який, як очікується, деформує наявні області стійкості та змінює відповідні нелінійно-дисперсійні механізми.
	
\keyu{модуляційна стійкість, міжфазні хвилі, двошарова рідина, нестійкість Бенджаміна–Фейра, поверхневий натяг}
	
\end{abstract}


\EndPaper


\end{document}